\documentclass[%
 aip,
 amsmath,amssymb,
preprint,
floatfix,
]{revtex4-1}

\usepackage{graphicx}
\usepackage{float}
\usepackage{comment}
\usepackage{bm}
\usepackage{hyperref}
\hypersetup{colorlinks=true,urlcolor=blue,citecolor=blue,linkcolor=blue}
\usepackage{xcolor}
\usepackage[utf8]{inputenc}
\usepackage[T1]{fontenc}
\usepackage{mathptmx}
\usepackage{etoolbox}
\def\bx{{\bf x}}
\def\bh{{\bf h}}
\def\ba{{\bf a}}

\def\bd{{\bf d}}
\def\bm{{\bf m}}

\makeatletter
\def\@email#1#2{%
 \endgroup
 \patchcmd{\titleblock@produce}
  {\frontmatter@RRAPformat}
  {\frontmatter@RRAPformat{\produce@RRAP{*#1\href{mailto:#2}{#2}}}\frontmatter@RRAPformat}
  {}{}
}%

\makeatother
\begin{document}

\title{ Assessing generative modeling approaches for free energy estimates in condensed matter } 

\author{Maximilian Schebek}
\affiliation{Department of Physics, Freie Universit\"at Berlin,
14195 Berlin, Germany}

\author{Jiajun He}
\affiliation{Department of Engineering, University of Cambridge, Cambridge CB2 1PZ, United Kingdom}

\author{Emil Hoffmann}
\affiliation{Department of Physics, Freie Universit\"at Berlin,
14195 Berlin, Germany}

\author{Yuanqi Du}
\affiliation{Department of Computer Science, Cornell University, 14850 Ithaca, NY, USA}

\author{Frank No\'e}
\affiliation{Department of Physics, Freie Universit\"at Berlin,
14195 Berlin, Germany}
\affiliation{Department of Mathematics and Computer Science, Freie Universit\"at Berlin, 14195 Berlin, Germany}
\affiliation{Microsoft Research AI for Science, 10178  Berlin, Germany}
\affiliation{Department of Chemistry, Rice University, Houston, Texas 77005, USA}

\author{Jutta Rogal}
\email{jrogal@flatironinstitute.org}
\affiliation{Initiative for Computational Catalysis, Flatiron Institute, New York, New York 10010, USA}

\date{\today}

\begin{abstract}
The accurate estimation of free energy differences between two states is a long-standing challenge in molecular simulations. Traditional approaches generally rely on sampling multiple intermediate states to ensure sufficient overlap in phase space and are, consequently, computationally expensive. Boltzmann Generators and related generative-model-based methods have recently addressed this challenge by learning a direct probability density transform between two states. However, it remains unclear which approach provides the best trade-off between efficiency, accuracy, and scalability. In this work, we review and benchmark selected generative approaches for condensed-matter systems, including discrete and continuous normalizing flows for targeted free energy perturbation and FEAT (Free Energy Estimators with Adaptive Transport) combined with the escorted Jarzynski equality, using coarse-grained monatomic ice and Lennard-Jones solids as benchmark systems. All models yield highly accurate free energy estimates and, depending on the system, may require fewer energy evaluations than traditional methods. Continuous flows and FEAT are most efficient in energy evaluations, whereas discrete flows have substantially lower inference cost.
By releasing all data together with our results, we enable future benchmarking of free energy estimation methods in condensed-phase systems.

\end{abstract}

\maketitle

\section{Introduction}

Free energy calculations play a central role across the natural sciences, providing a fundamental link between microscopic interactions and macroscopic thermodynamic properties.~\cite{tuckerman2023statistical} In computational materials science,  accurate free energy estimates are crucial for determining the relative stability of different phases of matter, which is experimentally challenging and often leaves uncertainty as to whether the observed phases are truly the most stable or if additional, unobserved phases exist.~\cite{Chew2023-eo, Vega2008} Despite their importance, estimating free energies remains a significant challenge, as absolute free energies are intractable for all but the simplest systems, and computational methods are limited to estimating differences in free energy between states.

Free energy perturbation (FEP)~\cite{zwanzig_fep} and derived estimators such as Bennett’s acceptance ratio (BAR)~\cite{Bennett1976-ic} rely on importance sampling from equilibrium distributions, where one distribution serves as a proposal for another. While formally exact, this approach becomes unfeasible when the two states share little phase-space overlap, resulting in large variance. 
The multistate BAR (MBAR)~\cite{Shirts2008-mbar} approach addresses this issue by introducing a chain of overlapping intermediate distributions, bridging large differences between states at the cost of additional simulations that may not be physically informative. 
In targeted free energy perturbation (TFEP),~\cite{tfep} 
an invertible transformation is introduced that maps samples from one distribution to another, thereby eliminating the need for intermediate stages. However, the efficiency of TFEP depends critically on the quality of this transformation, which can be hard to define based on physical intuition.

Thermodynamic integration (TI)~\cite{Kirkwood1935-qx} follows a different strategy by constructing a continuous interpolation path between two systems and computing the free energy difference by integrating ensemble averages of the instantaneous energy derivative along this path. In practice, discretization of the path into intermediate system states is required, which again incurs significant computational cost.

While the methods described thus far rely on samples drawn from equilibrium distributions at each state, the Jarzynski equality~\cite{jarzynski1997nonequilibrium} provides an alternative framework by establishing a connection between the nonequilibrium work required to drive the system from one equilibrium state to another and the corresponding difference in equilibrium free energies. The escorted Jarzynski equality improves convergence by introducing a control term that guides configurations along the switching process, reducing dissipation.~\cite{vaikuntanathan2008escorted} However, similar to the challenge faced in TFEP, the efficiency of the method depends critically on identifying a suitable transformation.

Boltzmann Generators~\cite{Noe2019} and follow-up work have proposed to sample equilibrium states by leveraging  deep neural networks to transform a tractable prior or reference distribution to a complex equilibrium distribution. This approach permits both configurational and thermodynamic free energy differences to be constructed because the learned probability transform contains the information of nonequilibrium work between the reference state and the target equilibrium state.~\cite{Noe2019} Subsequent work has leveraged this idea to perform generative equilibrium sampling of molecular and condensed-phase systems and to construct increasingly robust and scalable TFEP estimators with discrete normalizing flows,~\cite{wirnsberger_lfep, schebek2024efficient, schebek2025scalable, Rizzi2023, Wirnsberger_2022, Olehnovics2025, schebek2025estimating, falkner2023conditioning, rehman2026efficient, coretti_learning} where the transformation is expressed as a finite sequence of invertible transformations, and continuous normalizing flows,~\cite{klein2023equivariant, Zhao2023, erdogan2025, Hoffmann2026Boltzmann} where the transformation arises from integrating neural ordinary differential equations.

Neural approaches for thermodynamic integration using diffusion models~\cite{ho2020ddpm} have been proposed,~\cite{mate2024neural,mate2025solvation} however, they suffer from a heavy reliance  on problem-specific preconditioning.~\cite{mate2024neural,he2025feat} 
Most recently, nonequilibrium formulations were developed employing neural network-based strategies to improve convergence of the escorted Jarzynski equality through a learnable control term, introduced as \emph{Free energy Estimators with Adaptive Transport} (FEAT).~\cite{he2025feat}

All these approaches have shown promise for efficient free energy estimation, yet it remains unclear  which method provides the best trade-off between data efficiency, accuracy, and scalability for a given application.~\cite{john2025comparison, Olehnovics2024} This is particularly true for challenging condensed-phase systems, which are less explored compared to molecular systems, as they require large system sizes to achieve thermodynamic convergence as well as a careful treatment of periodic boundary conditions.

In this work, we systematically compare free energy estimation frameworks based on Boltzmann Generators and related deep generative models, focusing on crystalline systems subject to periodic boundary conditions. We benchmark both discrete and continuous flow models combined with the TFEP estimator and FEAT together with the escorted Jarzynski equality. 
We assess efficiency in training and inference, accuracy of free energy estimates, data requirements, and scalability for two condensed phase systems using  monoatomic water (mW)~\cite{molinero2009water} and Lennard-Jones (LJ) potentials. For both systems, we evaluate absolute and relative free energies for different crystalline phases, cubic/hexagonal mW ice and face-centred cubic (FCC)/hexagonal closed-packed (HCP) LJ, and explore different system sizes. We place particular emphasis on the required number of energy evaluations and show that, depending on the system, generative models are competitive with or can even outperform traditional methods.

\section{Probabilistic models for free energy estimates}
The free energy of a system is directly related to its partition function. Within the canonical ($NVT$) ensemble, with fixed number of particles $N$, temperature $T$, and volume $V$, the equilibrium distribution is given by the  Boltzmann distribution~\cite{tuckerman2023statistical, Frenkel2001-yl}
\begin{align}\label{eq:boltzmann}
    p(\bx) = \frac{e^{-u(\bx)}}{Z}   \quad ,
\end{align}
where  \(u(\bx)=\beta U(\bx)\) is the reduced potential with the potential  energy \( U(\bx)\) of  configuration $\bx\in \mathbb{R}^{3N}$, Boltzmann's constant $k_B$ and $\beta= 1/k_BT$. From the configurational partition function, $Z = \int e^{-u(\bx)} d\bx$,  the configurational contribution to the  Helmholtz free energy \(  F \)  is defined as
\begin{equation}\label{eq:f_logz}
    F =  - k_BT\ln Z   \quad .
\end{equation}
The free energy further includes a momentum contribution, which can be evaluated analytically.~\cite{tuckerman2023statistical} All absolute free energies reported in the current study include the momentum contribution. 

As the partition function cannot be expressed as an ensemble average, it cannot be obtained from molecular dynamics (MD) or Monte Carlo (MC) simulations directly. A more accessible quantity is the difference in free energy between two states $A$ and $B$, $\Delta F_{AB}$, which can be estimated using a variety of methods. 
For absolute free energies, a reference state $A$ with a tractable partition function and, correspondingly, computable free energy $F_A$ can be chosen, such that $F_B = F_A + \Delta F_{AB}$.
For crystalline phase, the Einstein crystal~\cite{Frenkel1984, Hoover1968} is a commonly used reference system which consists of $N$ independent harmonic oscillators coupled to a reference lattice. Since the displacements from the reference lattice follow a Gaussian distribution with a defined mean and variance, independent samples can directly be drawn from this reference distribution. 
In the following, we first review classical free energy estimators and how recent neural network-based probabilistic models can be applied in this context, and then discuss practical training schemes for the different models.

\subsection{Free Energy Estimators}

\begin{figure}[t!]
    \centering
    \includegraphics[width=1.0\linewidth]{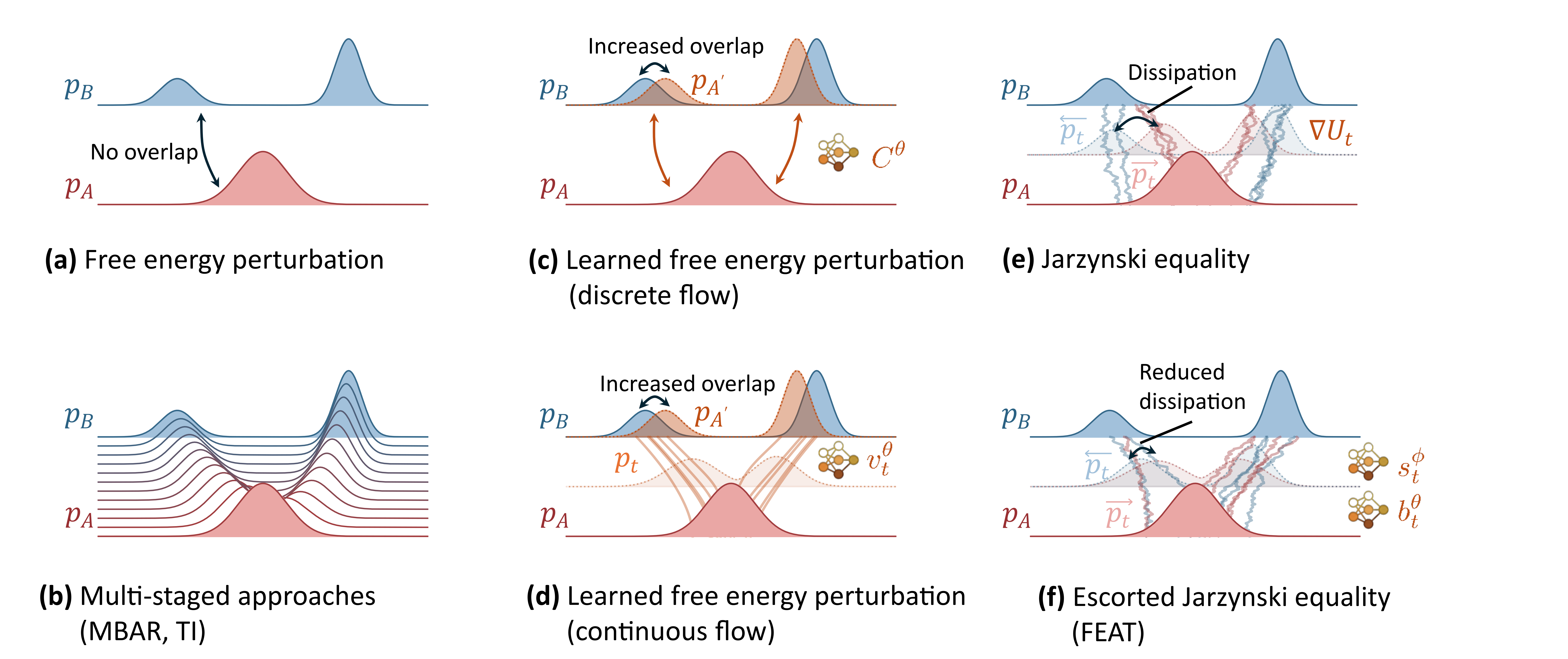}
    \caption{Overview of free energy estimation methods. $p_A$ and $p_B$ denote the equilibrium distributions of two systems $A$ and $B$. (a) Free energy perturbation (FEP) fails when two systems have no overlap. (b) Multi-staged approaches interpolate between $p_A$ and $p_B$  through a sequence of mutually overlapping intermediate distributions $p_i$ (MBAR, TI). (c) and (d) Learned FEP constructs an invertible map \(f\) to bridge the two non-overlapping distributions. Discrete flows learn a conditioner $C^\theta$ parameterizing a bijecting function, whereas continuous flows learn a time-dependent transport vector field \(v^\theta_t\). (e) Jarzynski equality utilizes the work along stochastic nonequilibrium trajectories with a time-dependent potential energy $U_t$. (f) Escorted Jarzynski equality introduces an additional control term $b_t$ to minimize the dissipation along the nonequilibrium trajectories and, in FEAT, both the score $s^\phi_t = \nabla u^\phi_t$ and control term $b^\theta_t$ can be learned.}
    \label{fig:fe_main2}
\end{figure}

A broad class of estimators is based on FEP,~\cite{zwanzig_fep} an importance-sampling-based estimator where samples from one state's equilibrium distribution is evaluated in both potentials
\begin{equation}\label{eq:f_diff}
     \Delta F_{AB} = - k_B T \ln  \mathbb{E}_{\bx\sim p_A} \biggl[\exp\Bigl(- \left(u_{B}(\bx) - u_A(\bx)\right)\Bigr)\biggr] \quad.
\end{equation}
While FEP has the advantage of only requiring samples from one of the two states, it shows high variance when the two states have limited overlap (Fig.~\ref{fig:fe_main2}(a)), which is especially problematic in high-dimensional systems. To reduce variance, BAR~\cite{Bennett1976-ic} extends FEP to provide a minimum-variance estimator using samples from both states, though it still requires sufficient overlap between the two distributions. MBAR~\cite{Shirts2008-mbar} further addresses this limitation by introducing multiple intermediate states, improving convergence at the cost of increased computational expense, illustrated in Fig.~\ref{fig:fe_main2}(b).

Alternatively, the problem can be approached by constructing a bridge between the two distributions 
and leveraging similar importance-sampling-based estimators over the bridge to compute free energy differences. Different realizations of this bridge lead to different free energy estimators. 

One way to realize this bridge has been proposed in targeted FEP~\cite{tfep} where an invertible and deterministic map
$f:\mathbb{R}^{3N}\rightarrow \mathbb{R}^{3N}$ is introduced
that transforms samples $\bx$ from a distribution $p_A$  to  $\bx' = f(\bx)$ with $\bx' \sim p_{A'}$, and where $p_{A'}$ has increased overlap with the target distribution $p_B$ (Fig.~\ref{fig:fe_main2}(c) and~(d)).
The probability of the transformed sample is given by the change-of-variable formula
\begin{equation}\label{eq:chofvariable}
    p_{A'}(\bx') = p_A(\bx) \left| \det \frac{\partial f(\bx)}{\partial \bx}\right|^{-1} \quad ,
\end{equation}
and the corresponding free energy difference is calculated through importance weights with 
\begin{equation}\label{eq:tfep_disc}
    \Delta F_{AB} = - k_B T \ln \mathbb{E}_{\bx\sim p_A}\biggl[\exp\Bigl(-(u_B(f(\bx))-u_A(\bx) - \ln |\det \frac{\partial f(\bx)}{\partial \bx}|)\Bigr)\biggr] \quad .
\end{equation}
Defining a suitable map $f$ by hand is, however, a non-trivial task. As discussed in the following section, the map can be naturally realized via normalizing flows,~\cite{papamakarios_flow, lipman2022flow}  neural-network based bijections between probability distributions, which was first proposed in the framework of Boltzmann Generators and the related~\cite{Noe2019} learned free energy perturbation (LFEP).~\cite{wirnsberger_lfep} 

By including concepts from nonequilibrium thermodynamics, 
the bridge can further be defined using stochastic processes. The Jarzynski equality~\cite{jarzynski1997nonequilibrium} relates the  nonequilibrium work $W$ to the equilibrium free energy difference
\begin{equation}
\Delta F_{AB} = -k_B T \ln \mathbb{E}_\mathbb{\overrightarrow{\mathbb{P}}}\left[ \exp(-  w)\right] \quad, 
\qquad w = \int_0^1 \partial_t u_t(\bx_t) \text{d}t \quad, 
\end{equation}
where  $w = \beta W$ denotes the reduced nonequilibrium work accumulated along the path with the time-dependent reduced potential  $u_t = \beta U_t$ with $u_0 = u_A$ and $u_1 = u_B$,  
and $\overrightarrow{\mathbb{P}}$ is the path measure of the solution of the overdamped Langevin equation
\begin{equation} \label{eq:langevin}
    \text{d} \bx_t = -\sigma_t^2 \nabla U_t(\bx_t) \text{d}t + \sqrt{2}\sigma_t \text{d}\overrightarrow{B_t}, \quad \bx_0\sim p_A \quad .
\end{equation}
$B_t$ is the Brownian motion where the arrow indicates the forward direction, and $\sigma_t \geq 0$ is related to the diffusion coefficient.
 The Jarzynski equality can also be understood as an importance-sampling-based approach over path space 
and to reduce variance, samples from forward and backward paths can be used, Fig.~\ref{fig:fe_main2}(e). 
The corresponding, more general relationship is provided by Crooks fluctuation theorem~\cite{crooks1999entropy}
\begin{equation} \label{eq:crooks}
\frac{\text{d}\overleftarrow{\mathbb{P}}} {\text{d}\overrightarrow{\mathbb{P}}} = \exp(- w + \beta \Delta F_{AB})  \quad ,
\end{equation}
where $\overleftarrow{\mathbb{P}}$ denotes the path measure of the backward stochastic process.
It is worth noting that Crooks fluctuation theorem also provides an additional way to calculate the work by directly evaluating the Radon-Nikodym derivative (RND) in the path space, $\text{d}\overleftarrow{\mathbb{P}} / \text{d}\overrightarrow{\mathbb{P}}$.~\cite{he2025feat}

In practice, the stochastic processes considered in the Jarzynski equality can result in large dissipation, an issue which can be mitigated by introducing an additional control term $b_t$~\cite{vaikuntanathan2008escorted} in the corresponding stochastic differential equation (SDE)
\begin{equation} \label{eq:fwd_sde}
    \text{d} \bx_t = -\sigma_t^2 \nabla U_t(\bx_t) \text{d}t + b_t(\bx_t) \text{d}t + \sqrt{2}\sigma_t \text{d}\overrightarrow{B_t}, \quad \bx_0\sim p_A \quad .
\end{equation}
This yields the escorted Jarzynski equality
\begin{equation}\label{eq:esc_jarzinsky}
\Delta F_{AB} = -k_B T\ln \mathbb{E}_\mathbb{\overrightarrow{\mathbb{P}^b}}\left[ \exp(-\widetilde{w})\right]  \quad ,
\end{equation}
where $\overrightarrow{\mathbb{P}^b}$ denotes the path measure of the stochastic process with additional $b_t$, Fig.~\ref{fig:fe_main2}(f), and  $\widetilde{w}$ is the reduced generalized work 
\begin{align}\label{eq:esc_jarzinsky_work}
\widetilde{w} = \int_0^1 \big(\partial_t u_t(\bx_t) + b_t(\bx_t) \cdot \nabla u_t(\bx_t) + \nabla \cdot b_t(\bx_t)\big)\text{d}t \quad .
\end{align}
Note that this equation holds for any $\sigma_t \geq 0$.

Recently, Crooks fluctuation theorem was generalized in this setting by introducing the additional control term in both the forward and backward stochastic processes~\cite{vargastransport,zhong2024time}
\begin{equation} \label{eq:esc_crooks}
\frac{\text{d}\overleftarrow{\mathbb{P}^b}} {\text{d}\overrightarrow{\mathbb{P}^b}} = \exp(-\widetilde{w} + \beta \Delta F_{AB}) \quad ,
\end{equation}
where $\overleftarrow{\mathbb{P}^b}$ denotes the path measure of the backward SDE with additional $b_t$
\begin{equation} \label{eq:bwd_sde}
    \text{d} \bx_t = \sigma_t^2 \nabla U_t(\bx_t) \text{d}t + b_t(\bx_t) \text{d}t + \sqrt{2}\sigma_t \text{d}\overleftarrow{B_t}, \quad \bx_1\sim p_B \quad .
\end{equation}
Similar to Eq.~\eqref{eq:crooks}, Eq.~\eqref{eq:esc_crooks} allows us to view the escorted Jarzynski equality as an importance-sampling-based approach in path space, 
and to compute $\widetilde{w}$ directly from the path RND.~\cite{he2025feat} As for TFEP, defining suitable $u_t$ and $b_t$ by hand is a challenging task.  However, parameterizing them as neural networks provides a flexible and expressive framework for learning nonequilibrium free energy estimators, which is realized in FEAT.~\cite{he2025feat}

The extra control term $b_t$ enables the distribution of the samples along the transport to better align with the adiabatic path between $u_A$ and $u_B$.
In the limit of no dissipation, the transport is fully adiabatic and the system stays in equilibrium along the entire path. This is utilized in thermodynamic integration
\begin{equation} \label{eq:TI}
\Delta F = \int_0^1 \mathbb{E}_{\bx\sim p_t} [ \partial_t U_t (\bx)] \text{d}t \quad ,
\end{equation}
where obtaining the energy interpolant $U_t$ and samples $\bx \sim p_t$ can be made more efficient with neural network parameterized transport models, known as Neural~TI.~\cite{mate2024neural,mate2025solvation}

\subsection{Training  and Inference Strategies for Neural Free Energy Estimators} 
The neural free energy estimators based on generative ML models assessed within this work, discrete and continuous normalizing flows together with TFEP and FEAT together with escorted Jarzynski, differ in their training objectives and associated free energy evaluations.
While discrete normalizing flows can be trained having access only to samples from the prior distribution and the unnormalized density of the target (energy-based training), both CNFs and FEAT require samples from the prior and target distribution. The corresponding training objectives are discussed below, where trainable quantities are indicated by superscripts $\theta$ and $\phi$.

\paragraph{Discrete normalizing flows.} 
Discrete normalizing flows (DNFs) are commonly implemented based on coupling layers,~\cite{dinh_nice_2014,dinh2017density} which partition the input into two channels \(\bx = (\bx_{\alpha}, \bx_{\beta})\). One channel is transformed by a bijector \(g\) according to \(\bx_\alpha' = g(\bx_{\alpha} \mid C^{\theta}(\bx_{\beta}))\), where the parameters of \(g\) are produced by a conditioner \(C^{\theta}\) acting on the other channel.  This construction yields a Jacobian that is analytically expressible in terms of the bijector parameters, and whose triangular structure allows its determinant to be computed easily. The full transformation is a composition of $K$ coupling layers \(f^{\theta}(\bx) = g_{K-1} \circ\cdots \circ g_0\) with changing partitioning of the channels. 
The exact and extremely efficient density estimation enables to train discrete normalizing flows by minimizing the reverse Kullback-Leibler (KL) divergence  between the generated distribution $p_{\theta}$ and the target distribution $p_B$~\cite{Noe2019, wirnsberger_lfep}
\begin{equation}\label{eq:ene_loss}
\mathcal{L}_{\text{reverse KL}}(\theta) = D_{\text{KL}}(p_{\theta} \|p_B) 
\propto \mathbb{E}_{\bx\sim p_A}\Bigl[-u_A(\bx) - \ln |\det \frac{\partial f^\theta(\bx)}{\partial \bx}|
+ u_B(f^\theta(\bx))  \Bigr]\quad ,
\end{equation}
where the generated samples $\bx' = f^{\theta}(\bx)$ are distributed according to $\bx' \sim p_{\theta}$.
In TFEP, the ideal mapping is given if the generated distribution is equal to the target distribution, that is $D_{\text{KL}}(p_{\theta} \|p_B) = 0$, and  $\Delta F_{AB}$ provides a lower bound to the loss function in Eq.~\eqref{eq:ene_loss}, correspondingly.
In our simulations, the prior distribution $p_A$ is represented by the Einstein crystal,~\cite{Frenkel1984} and samples $\bx \sim p_A$ are readily available from the corresponding Gaussian distribution. Crucially, no samples of the target distribution are required for reverse KL training. To evaluate Eq.~\eqref{eq:ene_loss}, only the energy of the generated sample in the target potential, $u_B(\bx')$, is required as well as the determinant of the Jacobian of $f^{\theta}$. 
After training, the TFEP estimator in  Eq.~\eqref{eq:tfep_disc} can directly be used together with the learned invertible mapping $f^{\theta}$ to estimate free energy difference $\Delta F_{AB}$.

\paragraph{Continuous normalizing flows.}
Similarly, the TFEP bridge can be formulated as an ordinary differential equation (ODE)~\cite{chen2018neural} with a learnable time-dependent vector field $v_t^{\theta}: \mathbb{R}^{3N}\times [0, 1] \rightarrow \mathbb{R}^{3N}$ from state $A$ to $B$, Fig.~\ref{fig:fe_main2}(d).
In contrast to coupling flows, evaluating the density induced by continuous normalizing flows (CNFs) is computationally expensive, as it requires integrating an ODE and, in particular, computing the divergence of the vector field. This cost prevents CNFs from being efficiently trained via reverse KL minimization. Instead, learning the vector field of the corresponding ODE connecting samples from the prior and target distributions can be achieved with the conditional flow matching (CFM)~\cite{lipman2022flow} objective which 
requires access to samples from both states $A$ and $B$, but no density evaluation. 
While samples from the Gaussian prior $p_A$ are directly available, uncorrelated samples from the target distribution $p_B$ need to be generated with MD or MC simulations.
The corresponding loss function is
\begin{equation}\label{eq:fm_loss}
\mathcal{L}_{\text{CFM}}(\theta) = \mathbb{E}_{(\bx_A, \bx_B)\sim \pi, t\sim \mathcal{U}(0, 1)} \Bigl[ | v_t^\theta(\bx_t) - \dot{\bx}_t|^2 \Bigr]
\quad ,
\end{equation}
where $\bx_t$ is given by an interpolant function $\bx_t = I_t(\bx_A,\bx_B) = \alpha_t \bx_A + \beta_t \bx_B$ with boundary conditions $\alpha_0 = 1, \alpha_1 = 0$ and $\beta_0 = 0, \beta_1=1$, and $\pi \in  \Pi(p_A, p_B)$ denotes a coupling of the two distributions with marginals $p_A$ and $p_B$.
A common choice for $I_t$ is a linear interpolation $I_t = (1-t)\bx_A + t\bx_B$, and common couplings include independent samples, $\bx_A \sim p_a$ and $\bx_B \sim p_B$, or the optimal transport (OT) plan of $(\bx_A,\bx_B)$-pairs.~\cite{tong2023improving}
$\dot{\bx_t} = \partial_t \bx_t=\partial_t (\alpha_t \bx_A + \beta_t \bx_B)$ is the time derivative of the interpolant function. 
After training, the instantaneous change-of-variable formula with $v_t^\theta$ can be applied to estimate free energy differences by integrating the ODE to generate samples $\bx_1$ and evaluating the divergence $\nabla \cdot v_t^{\theta}(\bx_t)$ according to ~\cite{Zhao2023, chen2018neural}
\begin{equation}\label{eq:tfep_cont}
    \Delta F_{AB} = - k_B T \ln \mathbb{E}_{\bx_0\sim p_A}\bigl[\exp\big(-(u_B(\bx_1)-u_A(\bx_0) -  \int_0^1 \nabla \cdot v^{\theta}_t(\bx_t)\text{d}t \big)\bigr] \quad.
\end{equation}

The exact computation of the divergence quickly becomes prohibitively expensive, requiring approximations such as  Hutchinson's estimator.~\cite{Skilling1989TheEO,Hutchinson01011990} 
For condensed phase systems with periodic boundary conditions, the distributions do not lie on the Euclidian manifold but on $3N$-dimensional tori and the interpolant as well as divergence in Eq.~\eqref{eq:tfep_cont} need to be computed correspondingly.~\cite{chen2024flowmatchinggeneralgeometries}

\paragraph{Escorted Jarzynski equality with FEAT.} 
The escorted Jarzynski estimator requires to evaluate a time-dependent potential energy function $u_t$  and  control term $b_t$. 
One way to realize this is to predefine a fixed energy interpolant $u_t$ while learning the control term $b_t$ to reduce the mismatch between the distribution of transported data $\bx_t$ and the energy $u_t$.~\cite{vargastransport,zhong2024time,albergonets} 
However, this training strategy can suffer from low efficiency~\cite{he2025no} and its stability strongly depends on the chosen energy interpolation.~\citep{mate2023learning}
By contrast, FEAT~\cite{he2025feat} learns both the time-dependent energy $u_t$ (in the form of the score $s_t = \nabla u_t$) and the control term $b_t$ directly from the data.
Known as the stochastic interpolant objective,~\cite{albergo2025stochastic} FEAT does not require to simulate trajectories during training 
improving scalability.

The loss functions for the control term $b_t^{\theta}$ and score $s_t^{\phi}$ are given by CFM and denoising score matching (DSM),~\cite{vincent2011connection} respectively
\begin{align}
\mathcal{L}_{\text{CFM}} (\theta) &= \mathbb{E}_{(\bx_A, \bx_B)\sim \pi, t\sim \mathcal{U}(0, 1),  \epsilon \sim \mathcal{N}(0, \text{Id})} \Bigl[ | b_t^\theta(\bx_t) - \dot{\bx}_t|^2 \Bigr] \label{eq:feat_loss1}\\
\mathcal{L}_{\text{DSM}}(\phi) &= \mathbb{E}_{(\bx_A, \bx_B)\sim \pi, t\sim \mathcal{U}(0, 1), \epsilon \sim \mathcal{N}(0, \text{Id})} \Bigl[ | s_t^\phi(\bx_t) - \gamma_t^{-1} \epsilon|^2 \Bigr] \label{eq:feat_loss2}  \quad ,
\end{align}
where  $\bx_t = I_t(\bx_A, \bx_B, \epsilon) = \alpha_t \bx_a + \beta_t \bx_b + \gamma_t \epsilon$, with boundary conditions $\alpha_0 = 1, \alpha_1 = 0$, $\beta_0 = 0, \beta_1=1$, and $\gamma_0=\gamma_1 = 0$, and $\dot{\bx}_t = \partial_t\bx_t$ is the corresponding time derivative. 
The total loss function is given by the sum of these two parts, $\mathcal{L}_{\text{FEAT}}(\theta,\phi) = \mathcal{L}_{\text{CFM}} (\theta) + \mathcal{L}_{\text{DSM}}(\phi)$.

If one of the distributions is Gaussian (as is the case for our Einstein crystal prior distribution $p_A$) and the coupling $\pi$ in Eqs.~\eqref{eq:feat_loss1} and~\eqref{eq:feat_loss2} leaves the distributions independent, i.e., $\pi = p_A \otimes p_B$, the score function and the control term are analytically related (see SI for details).~\cite{albergo2025stochastic}
Consequently, in our experiments we only learn the control term $b_t^\theta$, and compute the corresponding score function analytically.
Additionally, in this setting, the auxiliary random variable $\epsilon$ can be absorbed into the Gaussian component of the prior distribution, allowing us to use the same interpolant, $\bx_t = I_t(\bx_A, \bx_B) = \alpha_t \bx_A + \beta_t \bx_B$, as in the CNF case. 

After learning the control term $b_t^\theta$ (and deriving the corresponding score $s_t$ from it), the free energy difference can be estimated with the escorted Jarzynski equality in Eq.~\eqref{eq:esc_jarzinsky} by generating samples with the forward SDE in Eq.~\eqref{eq:fwd_sde} and evaluating the reduced work $\widetilde{w}$ along the corresponding trajectories.
However, the work defined in Eq.~\eqref{eq:esc_jarzinsky_work} requires computing a computationally expensive divergence and learning the energy function (instead of the score), which is a more difficult training task.
Instead, we calculate the work with Crooks fluctuation theorem generalized to the escorted case (Eq.~\eqref{eq:esc_crooks}) with 
\begin{align}
    \widetilde{w} &=  -\ln \frac{\text{d}\overleftarrow{\mathbb{P}^b}} {\text{d}\overrightarrow{\mathbb{P}^b}}+ \beta \Delta F_{AB}\\
      &= -\ln \frac{p_B(\bx_1)} {p_A(\bx_0)} + R + \beta\Delta F_{AB}\\ 
      &= u_B(\bx_1) - u_A(\bx_0)  +  R + \beta(F_A - F_B) + \beta\Delta F_{AB}\\ 
      &= u_B(\bx_1) - u_A(\bx_0)  + R\label{eq:work_fb_rnd}\quad,
\end{align}
where $R$ is the forward-backward path integral,~\cite{vargastransport,berner2025discrete,he2025rne} 
which can be calculated with It\^o's integral
\begin{multline} \label{eq:ito}
      R = -  \int_0^1 \frac{1}{2\sigma_t^2} (b_t^\theta(\bx_t) -  \sigma_t^2 s_t^\phi(\bx_t) )\cdot \text{d}  \overleftarrow{\bx_t} + \int_0^1 \frac{1}{2\sigma_t^2} (b_t^\theta(\bx_t) +  \sigma_t^2 s_t^\phi(\bx_t) )\cdot \text{d}  \overrightarrow{\bx_t}   \\  - \frac{1}{2}\int_0^1 \frac{1}{2\sigma_t^2} (\|b_t^\theta(\bx_t) +  \sigma_t^2 s_t^\phi(\bx_t) \|^2 - \|b_t^\theta(\bx_t) -   \sigma_t^2 s_t^\phi(\bx_t) \|^2)  \text{d}t \quad.
\end{multline} 
The arrows over $\bx_t$ indicate It\^o's forward or backward integral, respectively. 
This is  equivalent to the work calculated by Eq.~\eqref{eq:esc_jarzinsky_work} if the learned score perfectly matches the gradient in the two states, $s_0^\phi = \nabla u_A$ and $s_1^\phi = \nabla u_B$, respectively.
Numerically, the forward and backward SDEs are integrated using the Euler-Maruyama (EM) method and Eq.~\eqref{eq:ito} is discretized into a sum of log-density of Gaussian kernels (see SI for details).

\subsection{One-sided vs. Two-sided Estimators}
 If during inference samples from both distributions, the prior $p_A$ and target $p_B$, are available, the accuracy of the free energy calculation can be improved by constructing a minimum-variance estimator following the BAR principle.~\cite{Bennett1976-ic}
Taking FEAT as an example, one can not only simulate the forward SDE (Eq.~\eqref{eq:fwd_sde}) 
and compute the work along the forward trajectories, but also simulate the corresponding backward SDE (Eq.~\eqref{eq:bwd_sde})
and compute the work along the backward trajectories using the path RND in Eq.~\eqref{eq:work_fb_rnd}.
For clarity, we denote by $\widetilde{w}(\overrightarrow{\bx})$ and $\widetilde{w}(\overleftarrow{\bx})$ the work associated with the forward and backward trajectories $\overrightarrow{\bx}\sim \overrightarrow{\mathbb{P}}^b$ and $\overleftarrow{\bx} \sim \overleftarrow{\mathbb{P}}^b$, respectively.
The corresponding free energy estimator is
\begin{align}\label{eq:two-sided-F}
    \Delta F_{AB} = k_BT \ln 
    \frac{ \mathbb{E}_{\overleftarrow{\bx} \sim \overleftarrow{\mathbb{P}}^b}\bigl[\varphi\bigl( - \widetilde{w}(\overleftarrow{\bx}) + c \bigr) \bigr] }
    {\mathbb{E}_{\overrightarrow{\bx} \sim \overrightarrow{\mathbb{P}}^b}\bigl[\varphi\bigl(  \widetilde{w}(\overrightarrow{\bx}) - c \bigr) \bigr] } + c\quad,
\end{align}
where $\varphi(x) = \bigl(1+\exp(x)\bigr)^{-1}$ is the Fermi function and $c$ is an arbitrary constant, with the optimal choice given by $c = \Delta F_{AB}$. 
In practice,  the numerator and denominator are estimated with the same number of forward and backward trajectory samples, and the constant $c$ is randomly initialized and iteratively updated using the current estimate of $\Delta F_{AB}$
until convergence where $c \approx \Delta F_{AB}$. 

\section{Free energies of condensed phase systems}
\subsection{Experiments}

\begin{figure}[t!]
    \centering
    \includegraphics[width=.5\linewidth]{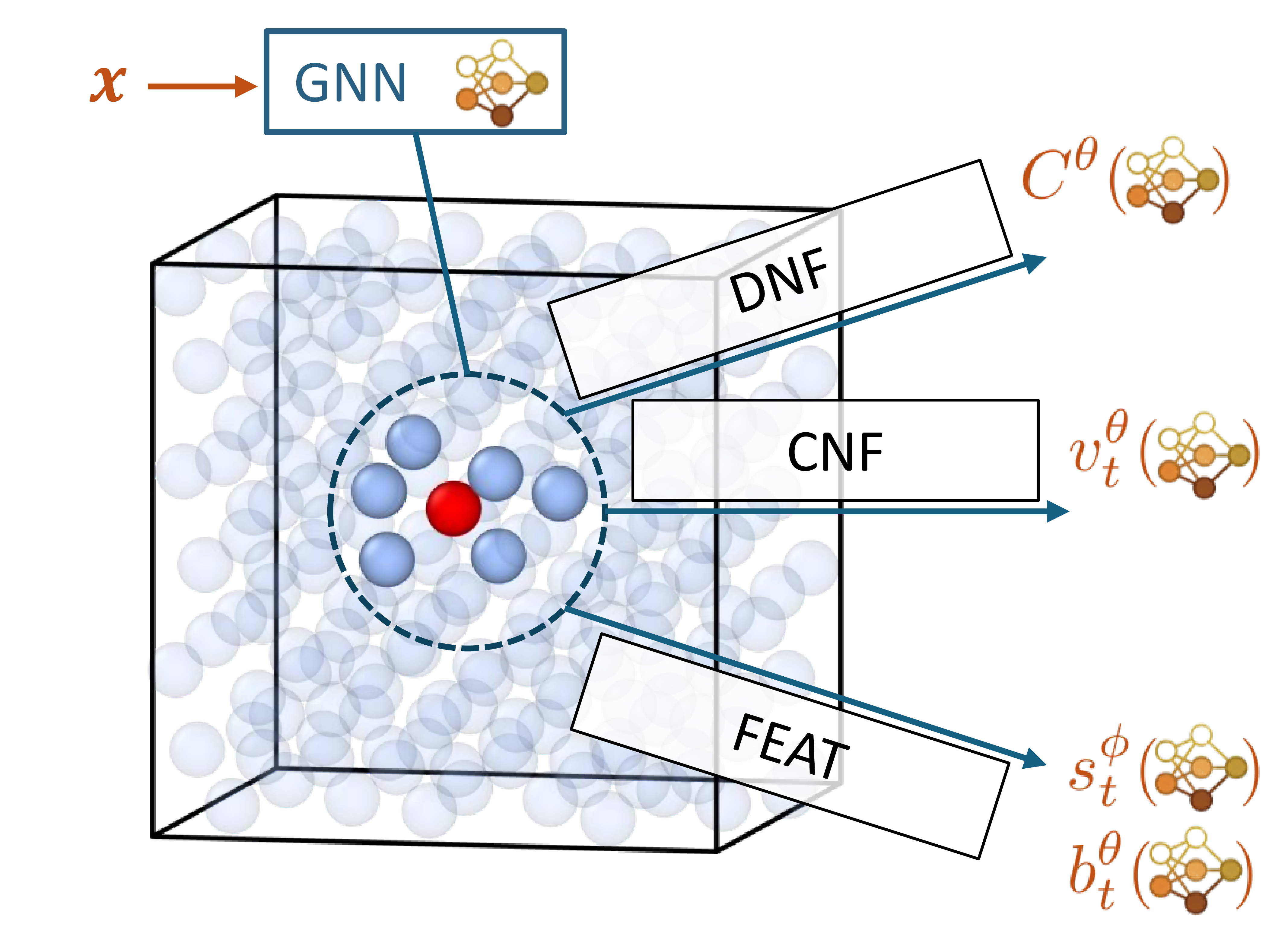}
    \caption{Overview of the featurization used for the different models. In all cases, a GNN encodes environment-dependent information by aggregating data from each particle’s neighborhood. For discrete coupling flows, this information defines the bijector function through the conditioner. For CNFs, it determines the vector field, and for FEAT the control term and score. 
    \label{fig:gnn}}
\end{figure}

We assess the performance of the generative models for free energy estimation discussed in the previous section on representative condensed-phase systems and evaluate their accuracy and efficiency for free energy evaluations.
As benchmark systems, we employ Lennard-Jones solids of varying sizes in the FCC and HCP crystalline structures, as well as the mW water model,~\cite{molinero2009water} for which we consider both the hexagonal and cubic ice phases. 
All simulations are performed at constant volume and temperature, details regarding the systems are given in the SI.
To enable a fair comparison, we represent the learnable components of all three approaches in a unified manner based on graph neural networks (GNNs),~\cite{gilmer2017neural} as they provide a natural and effective representation for condensed-phase systems. To model the vector field of CNFs as well as the control term and score in FEAT (see Fig.~\ref{fig:gnn}), we employ identical equivariant GNNs (EGNNs),~\cite{satorras2021equivariant, klein2023equivariant} ensuring that any performance differences between these two methods stem from their algorithmic design rather than architectural choices. We note that using a more expressive architecture, such as the equivariant transformer~\cite{pelaez_torchmdnet_2024} considered in Ref.~\onlinecite{Hoffmann2026Boltzmann}, may further improve the performance of CNFs and FEAT, but we deliberately do not adopt it here in order to maintain a controlled and directly comparable architectural setting across all three approaches. For  DNFs, which we implement as coupling flows, we use a non-equivariant but structurally similar GNN to parametrize the conditioner, based on a recently developed augmented flow architecture.~\cite{schebek2025scalable}  Since atomic motion in solids is largely confined to small oscillations around equilibrium lattice positions, the absence of rotationally equivariant structure in the coupling flows is unlikely to be a limiting factor. CNFs can be formulated directly on the appropriate manifold to naturally incorporate periodic boundary conditions,~\cite{chen2024flow, Hoffmann2026Boltzmann} whereas for DNFs and FEAT we instead model the particle displacements from the equilibrium lattice.~\cite{schebek2025scalable} To avoid biases in the free energy estimates arising from the use of Hutchinson probe vectors, we chose their number sufficiently large to ensure convergence of the CNF results.~\cite{Hoffmann2026Boltzmann} In all cases, the prior distribution is represented by an Einstein crystal.~\cite{Frenkel1984} Further details regarding the employed architectures and prior distributions are given in the SI.

To evaluate the computational efficiency of the different approaches, we analyze their performance as a function of the computational budget, measured by the  number of target energy evaluations required for model training and free energy estimation. The continuous models (FEAT and CNFs) rely on samples from the target distribution for training and are therefore optimized using data-driven objectives (see Eqs.~\eqref{eq:fm_loss}, \eqref{eq:feat_loss1} and~\eqref{eq:feat_loss2}). For these models, the training budget is controlled by limiting the number of target-distribution samples. 
In contrast, the discrete-time coupling flows are trained using an energy-based objective (see Eq.~\eqref{eq:ene_loss}) which does not require pre-generated target samples. For the coupling flows, the training budget is regulated by restricting the number of training steps, and thus the total number of energy evaluations. To systematically explore performance, we define three budget levels for both training and inference. The low, medium, and high training budgets correspond to $10^3$, $10^4$, and $10^5$ samples from the target distributions. This is  equivalent to $10^6$, $10^7$, and $10^8$ energy evaluations of the target potential for energy-based training, as generating training data using MD requires around 1,000 steps to decorrelate consecutive samples. For inference, the budgets are defined by $10^2$, $10^3$, and $10^4$ samples for computing the final free energy estimates. Due to the high cost of evaluating free energy estimates using CNFs, we restrict ourselves to three estimates per reported data point, obtained by evaluating each of three independently trained models once. 
Consequently, the limited number of model evaluations constrains the statistical resolution of the uncertainty estimates for the free energy and the ESS, particularly in larger systems with lower sampling efficiency.

\begin{figure}[ht!]
    \centering
        \includegraphics{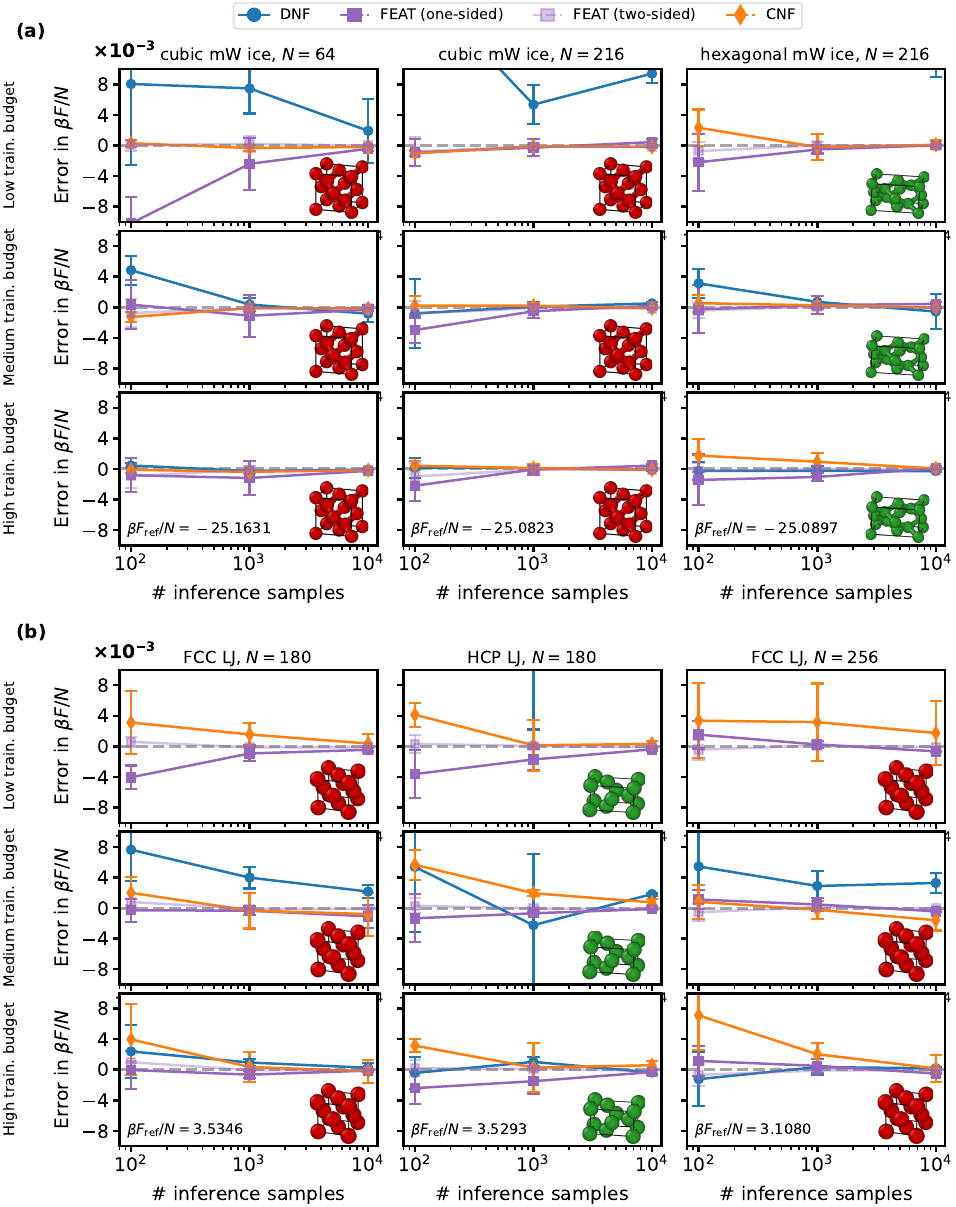}
    \caption{Error in the absolute free energies for (a) cubic and hexagonal mW ice and (b) FCC and HCP LJ as obtained from TFEP using discrete coupling flows and continuous flows, and from escorted Jarzynski equality using FEAT. Uncertainties were obtained by evaluating each of three independently trained models once.}
    \label{fig:results_main}
\end{figure}

In Fig.~\ref{fig:results_main}(a), the performance of the three approaches is compared for free energy estimation across three monatomic water systems, corresponding to  cubic (64 and 216 particles) and hexagonal ice (216 particles). For the flow-based models, free energies were obtained using the TFEP estimator (Eqs.~\eqref{eq:tfep_disc} and~\eqref{eq:tfep_cont}), while for FEAT we use both the one- and two-sided escorted Jarzynski equality (Eqs.~\eqref{eq:esc_jarzinsky} and ~\eqref{eq:two-sided-F}). With high training  and evaluation budgets (bottom row in Fig.~\ref{fig:results_main}(a)), all three models show excellent agreement with the reference absolute free energies, with deviations from the reference values significantly smaller than $10^{-3}k_BT$ per particle for both small and large systems (relative errors $\leq 4\cdot 10^{-5}$). For the cubic systems, both flow-based models already reach this level of accuracy with as few as 100 inference samples, whereas FEAT requires slightly more samples, although the two-sided estimator improves upon the one-sided version. 
It should be kept in mind, though, that the two-sided estimator requires additional uncorrelated samples from the target distribution during inference, incurring an additional cost of 1,000 energy evaluations per sample.
The high performance of the flow  models in the high training-budget regime is also reflected in the effective sample sizes (ESS)~\cite{Kish1965-es} (see SI for a definition) shown in Tab.~\ref{tab:ess}. The flow-based models achieve more than 60\% ESS for the 64-particle systems and over 15\% for the 216-particle systems.  Despite providing accurate free energy estimates, FEAT reaches only a few percent ESS for the large systems, which is however expected by design (see SI for a further comparison between FEAT and CNFs).
Curiously, the CNF results for the 216-particle hexagonal ice are somewhat worse than for the cubic phase which remains to be resolved. 

In the regime of medium training budget (middle row in Fig.~\ref{fig:results_main}(a)), corresponding to $10^7$ energy evaluations or $10^4$ training samples, CNFs and FEAT largely maintain their performance in terms of both bias and variance. The coupling flows are more strongly affected, which becomes particularly apparent when only 100 inference samples are used, leading to increased bias and variance. However, increasing the number of inference samples still yields highly accurate free energy estimates, comparable to those of the continuous models, except for the hexagonal 216-particle system, where a larger variance persists. These observations are also consistent with the ESS in Tab.~\ref{tab:ess}, which stays almost constant for FEAT and CNFs but dramatically decreases for the coupling flows.

The low training budget (top row in Fig.~\ref{fig:results_main}(a)) is clearly too small for the coupling flows and they perform poorly, especially for the larger systems. Even with a high inference budget of $10^4$ samples, they fail to recover the correct mean and exhibit very large variances. Interestingly, the continuous models remain capable of producing highly accurate free energy estimates, with one-sided FEAT performing well provided that a sufficient number of inference samples is used. CNFs consistently show strong performance across all settings, although for the hexagonal 216-particle system, a larger number of inference samples is required to sufficiently reduce the variance.  Here, coupling flows do not even reach one percent ESS, while CNFs and FEAT show only a slight decrease in ESS, see Tab.~\ref{tab:ess}.

Fig.~\ref{fig:results_main}(b) shows the corresponding results for three LJ systems: FCC structures with 180 and 256 particles, and HCP  with 180 particles. Similar to the findings for the mW system, coupling flows and FEAT achieve excellent agreement with reference calculations at high training and inference budgets. For the discrete coupling flows, performance is slightly worse for fewer inference samples compared to the mW systems 
but provides very good results with at least 1,000 samples.
As before, one-sided FEAT requires more inference samples, whereas the two-sided version converges at 1,000 samples. Interestingly, CNFs perform significantly worse for the LJ structures compared to the mW systems even at high training budgets: using 100 inference samples, CNFs show large deviations in the mean and high variance. Increasing the number of inference samples improves accuracy, although variance remains higher than in the corresponding mW case. Similar to mW, the performance of CNFs and FEAT at lower training budgets largely resembles that at high budgets, though one-sided FEAT requires more evaluation samples under these conditions. Coupling flows, however, degrade substantially at lower training budgets, becoming unreliable already at medium training levels. The poorer performance of the flow models is also reflected in the ESS (Tab.~\ref{tab:ess}). While FEAT achieves even higher ESS on the 180-particle LJ systems compared to the 216-particle mW systems, both continuous and discrete flow models lose substantial performance and now exhibit ESS values similar to FEAT.

\begin{figure}[t!]
    \centering
        \includegraphics{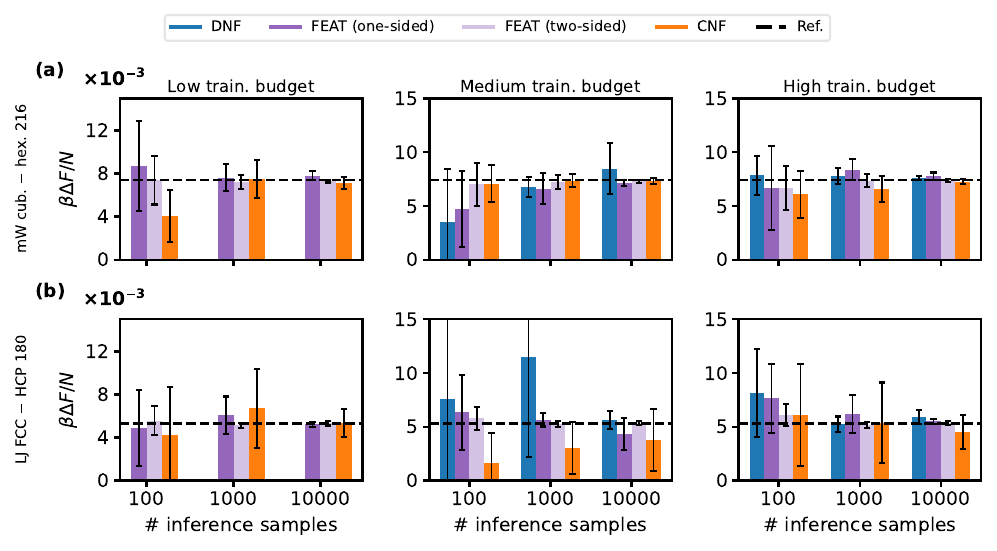}
    \caption{Free energy differences between (a) cubic--hexagonal mW ice and (b) FCC--HCP LJ phases as a function of the number of inference samples for low, medium, and high training budgets. The reference values are $\beta \Delta F_{\text{mW cub-hex}}/N = 0.0074$ and  $\beta \Delta F_{\text{LJ FCC-HCP}}/N = 0.0053$, respectively. Discrete-flow results for the low training budget are omitted due to large variance placing them outside the shown range.
    \label{fig:results_diffs}}
\end{figure}
Besides the evaluation of absolute free energies, a primary objective is the prediction of the relative stability of different phases which is given by their free energy difference. This is particularly challenging for the discussed systems, as their free energy differences are extremely small with $\beta \Delta F/N = 0.0074$ and $0.0053$ for mW cubic--hexagonal and LJ FCC--HCP phases, respectively.
The corresponding results are summarized in Fig.~\ref{fig:results_diffs} for 216-particle mW and 180-particle LJ systems.
Similar to our  observations for absolute free energies, for the  mW system (Fig.~\ref{fig:results_diffs}(a)), all methods accurately reproduce the reference results with extremely small error bars when trained with a high training budget (right graph) and evaluated with a large number of inference samples. At medium (middle graph) and low (left graph) training budgets, FEAT and CNFs still provide very good estimates, provided enough inference samples are used. Coupling flows deteriorate at lower training budgets: they are usable to a limited extent at medium training budgets when evaluated with many inference samples, even though the variance of the estimates can be reduced by evaluating the marginal generated density at the cost of additional inverse passes (see SI). For low training budgets, neither joint nor marginal estimates yield usable results and the values are not shown in the left graph of Fig.~\ref{fig:results_diffs}(a). For the LJ systems (Fig.~\ref{fig:results_diffs}(b)), both one- and two-sided FEAT remain highly accurate in predicting free energy difference even at low training budgets, provided the inference budget is large. At medium training budgets, CNFs perform similarly poorly as coupling flows, and while they improve at low training budgets, their variance remains very large, consistent with the trends observed in the absolute free energy calculations.

\begin{table*}[t!]
\centering
\caption{Effective sampling size in \%  evaluated using $10^4$ inference samples for cubic and hexagonal ice in the mW potential, and for FCC and HCP structures in the Lennard-Jones potential.\label{tab:ess}}

\setlength{\tabcolsep}{4pt}
\centering
\begin{tabular}{c|ccc|ccc|ccc}
\hline
\hline
\multicolumn{10}{c}{\textbf{mW systems}} \\
\hline
 & \multicolumn{3}{c|}{Low train. budget} & \multicolumn{3}{c|}{Medium train. budget} & \multicolumn{3}{c}{High train. budget} \\

\cline{2-10}
 & \makebox[1.3cm][c]{Cub64} & \makebox[1.3cm][c]{Cub216} & \makebox[1.3cm][c]{Hex216} & \makebox[1.3cm][c]{Cub64} & \makebox[1.3cm][c]{Cub216} & \makebox[1.3cm][c]{Hex216} & \makebox[1.3cm][c]{Cub64} & \makebox[1.3cm][c]{Cub216} & \makebox[1.3cm][c]{Hex216} \\
\hline
Coupling & 0.6(4) & 0.0(2) & 0.1(6) & 12.2(2) & 1.0(5) & 0.5(4) & 64.7(7) & 17.5(2) & 15.3(1) \\
CNF & 54.1(4) & 16.8(7) & 2.6(2) & 70.2(2) & 18.7(1) & 4.6(3) & 70.9(2) & 19.6(1) & 3.1(2) \\
FEAT (fwd) & 6.8(2) & 2.0(5) & 1.0(2) & 17.2(2) & 1.6(6) & 1.3(5) & 18.3(4) & 2.1(4) & 1.3(3) \\
FEAT (bwd) & 9.3(2) & 0.8(3) & 1.0(4) & 19.1(3) & 1.6(7) & 0.9(5) & 19.0(3) & 1.3(3) & 1.3(6) \\
\hline
\hline
\end{tabular}
\par
\vspace{1em}
\centering
\begin{tabular}{c|ccc|ccc|ccc}
\hline
\hline
\multicolumn{10}{c}{\textbf{LJ systems}} \\
\hline
 & \multicolumn{3}{c|}{Low train. budget} & \multicolumn{3}{c|}{Medium train. budget} & \multicolumn{3}{c}{High train. budget} \\
\cline{2-10}
 & \makebox[1.3cm][c]{FCC180} & \makebox[1.3cm][c]{FCC256} & \makebox[1.3cm][c]{HCP180} & \makebox[1.3cm][c]{FCC180} & \makebox[1.3cm][c]{FCC256} & \makebox[1.3cm][c]{HCP180} & \makebox[1.3cm][c]{FCC180} & \makebox[1.3cm][c]{FCC256} & \makebox[1.3cm][c]{HCP180} \\
\hline
Coupling & 0.1(3) & 0.0(7) & 0.0(8) & 0.3(2) & 0.1(1) & 0.1(1) & 2.6(1) & 2.7(6) & 2.8(2) \\
CNF & 0.5(3) & 1.3(2) & 0.3(5) & 0.7(8) & 3.1(5) & 0.6(4) & 0.6(3) & 4.0(7) & 0.9(3) \\
FEAT (fwd) & 1.4(6) & 0.5(2) & 2.2(9) & 1.9(1) & 0.7(1) & 2.5(2) & 2.5(8) & 0.5(1) & 2.4(1) \\
FEAT (bwd) & 2.4(3) & 1.2(3) & 2.6(2) & 2.2(10) & 1.0(5) & 2.2(1) & 2.1(8) & 0.9(4) & 2.4(2) \\
\hline
\hline
\end{tabular}
\vspace{0.5em}
\end{table*}

\subsection{Discussion}
\begin{figure}
    \centering
    \includegraphics[width=0.5\linewidth]{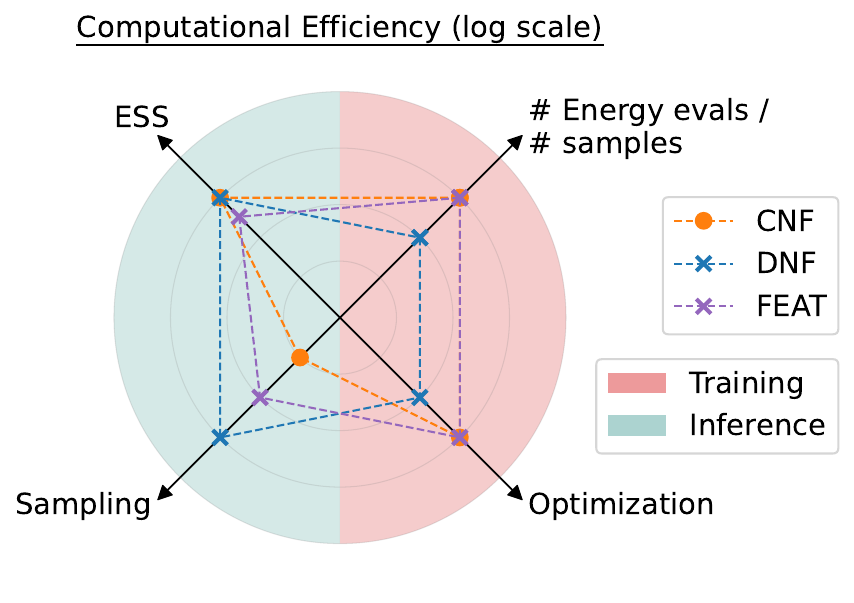}
    \caption{Efficiency comparison of three methods across four criteria: effective sample size (ESS), number of energy evaluations for training, optimization time, and time to compute a free energy estimate (sampling). Radial distance encodes relative efficiency for a given metric on a logarithmic scale. Larger values indicate greater efficiency. The background shading separates metrics relevant for inference (left) and training (right).}
    \label{fig:eff_comparison}
\end{figure}

While all three approaches are capable of accurately estimating free energy differences for our benchmark systems, they behave differently during training and inference. In particular, they differ substantially in the optimization time, the number of energy evaluations required to generate the training data for the continuous models or to train the coupling flows, the resulting ESS, and the sampling time required to compute a free energy estimate. 
A qualitative comparison of the computational efficiency across these four aspects is illustrated in Fig.~\ref{fig:eff_comparison}.

Remarkably, for the mW ice system, CNFs and FEAT yield highly accurate free energy estimates even when trained with only 1,000 samples. As detailed in the SI, both methods converge significantly faster than MD+MBAR in terms of the number of energy evaluations. For the Lennard-Jones system, in contrast, only FEAT converges faster than MD+MBAR, while CNFs show a similar convergence rate, consistent with the lower sampling efficiencies observed for this system. Coupling flows require around one order of magnitude more energy evaluations than the continuous models to reach similar performance, although for the ice system they still remain more efficient than MD+MBAR. Importantly, the continuous models are not only more efficient in terms of energy evaluations, but also in optimization time. Training the continuous models for $10^6$ steps takes about one day on a single GPU, compared with roughly three days on four GPUs for coupling flows, corresponding to approximately one order of magnitude reduction in optimization time. 
For the inexpensive  mW and Lennard–Jones model potentials used in this study, the combined time required to train and evaluate any of the three investigated models is significantly longer than that of a traditional MD+MBAR calculation, which typically takes only a few hours. 
However, for more costly interaction potentials, the wall clock time is increasingly dominated by calls to the energy function. In this regime, reducing the number of energy evaluations becomes the most important factor for overall efficiency.

The poorer performance of coupling flows at low training budgets likely stems from the limited number of gradient updates allowed by the energy-based training procedure. 
Replacing energy-based with data-based training using  samples from the target distribution would remove this limit on the number of training steps 
but, in practice, we found that this approach is prone to overfitting in the low-data regime. Continuous flows and FEAT are less affected by this limitation because their training objectives using data from prior and target states  continuously generates new samples from the easily sampled Gaussian prior, creating new $(\bx_A,\bx_B)$-pairs used for training.  This assumption might not hold anymore if these models were trained with a fixed number of samples from both the target and prior distributions.
If samples from the target are, however, not accessible with conventional sampling methods, energy-based training constitutes the only option, and recent approaches have proposed to formulate continuous models utilizing only the unnormalized density of the target samples.~\cite{vargas2023denoising,vargastransport} 
Nevertheless, energy-based training comes with several caveats. In particular, reverse KL minimization is known to be mode-seeking. Additionally, energy-based approaches often incur additional computational cost from backpropagating gradients through the simulation trajectories. In this scenario, with the chosen model architectures, the computational cost for training is highest for the CNFs, medium for FEAT, and less expensive for discrete coupling flows. 
A rigorous comparison is, however, difficult as it depends on the details of the architecture and hyperparameters.

Although coupling flows require longer training and more energy evaluations, they provide a crucial advantage during inference through direct density evaluation, making the computation of free energy estimates approximately one order of magnitude more efficient than FEAT and about two orders of magnitude more efficient than CNFs. In particular, estimating the free energy difference using 10,000 samples for the larger systems studied in this work takes less than one minute on a single NVIDIA A5000, while FEAT requires approximately 30 minutes and CNFs take around 5 hours even when using the Hutchinson's trace approximation.
This advantage becomes especially relevant for size-transferable models,~\cite{schebek2025scalable, Hoffmann2026Boltzmann} where generating samples for systems with more than a thousand atoms can become very expensive with continuous models. In such cases, coupling flows can effectively amortize the cost of the energy evaluations needed for training. 
Furthermore, while the evaluation of the exact Jacobian is analytical and, therefore, fast in coupling flows and not needed in FEAT, it becomes prohibitively expensive for larger system sizes with CNFs. While in this work we found that Hutchinson’s trace approximation appears to be sufficiently accurate, addressing this issue more generally remains an active area of research and is not yet entirely practical.~\cite{Peng2025,rehman2025falconfewstepaccuratelikelihoods,gloy2025hollowflowefficientsamplelikelihood}

For the systems investigated within this work, inference using one-sided estimators is generally sufficient, although more inference samples are required compared to a two-sided estimation as shown for FEAT. However, generating additional samples from the target for two-sided evaluations is costly in itself as every sample is roughly 1,000 times more expensive due to decorrelation in the MD sampling. Therefore, it can often be more efficient to increase the number of inference samples in the one-sided estimator instead of simulating samples from the target to be used in the two-sided estimator. Nevertheless, we note that for more challenging problems where the models exhibit poorer performance, two-sided estimators may still be required to obtain accurate results.

Scalability with system size  is  essential to reduce finite-size effects in condensed phase systems and can be a critical issue in generative models. In addition, the probability density is an extensive quantity, meaning a tighter per-atom convergence is needed to achieve comparable accuracy in total system quantities, as, for example, measured by the ESS. Already for the different system sizes investigated here, we see a clear decrease in performance for the different models for larger systems. Training models for larger systems becomes computationally expensive and convergence is much slower, such that for systems with more than a few hundred particles, these models are not yet fully applicable.
Some of us have recently proposed a size-transferable architecture based on augmented coupling layers that can be trained in small systems while inference with the trained model can be performed on much larger system sizes.~\cite{schebek2025scalable}
By representing the local environment around each particle with a GNN, similar ideas should hold for the continuous models, as has recently been shown for CNFs.~\cite{Hoffmann2026Boltzmann} While training costs and convergence is greatly improved in the small systems, inference must still be run with the entire large system. Here, CNF and FEAT like models might still be computationally challenging as inference costs scale roughly linear with system size. 

To fully amortize the training costs and make these models attractive for widespread applications, transferability not only in size but also between different systems would be desirable. One way to approach this is conditional training on atom types or external parameters.~\cite{falkner2023conditioning,schebek2024efficient,Moqvist2024}
Another idea is to train a general model and guide the generation of samples with specific properties during inference.~\cite{bao2023equivariantenergyguidedsdeinverse,Weiss2023guided}
Here, energy-based training has the advantage that samples under any condition can be evaluated whereas data-based training requires to have samples available over the full range of conditioning parameters.

It is also worth noting that, when comparing FEAT with deterministic mappings such as CNFs and coupling flows, the quality of the learned transport is not directly reflected by the ESS, as the importance sampling of the two approaches is performed in fundamentally different spaces. Deterministic mappings operate in the target (marginal) space, whereas FEAT performs importance sampling over the path space of the underlying stochastic differential equation. As a result, even for comparable transport quality, FEAT typically yields a lower ESS than CNFs or discrete coupling flows due to the data processing inequality, leading to higher estimator variance. In contrast, although CNFs often achieve higher ESS, this comes at the cost of increased computational overhead from evaluating the divergence term in the instantaneous change-of-variables formula  and requires a careful evaluation of the bias introduced by time discretization of the ODE and approximation of the Jacobian trace using Hutchinson's approximation. 
We provide a more detailed discussion of the trade-offs between variance, bias, and computational cost for FEAT and CNF in the SI.

While discrete and continuous models both have their advantages and disadvantages, the generally higher expressiveness of continuous architectures is likely to become important for more complex condensed phase systems, such as liquids or amorphous phases.  Improving their performance during inference for large-scale systems will thus be a crucial step in their applicability.

\section{Conclusion}
Overall, our results highlight the strengths of combining generative machine learning approaches with concepts from statistical mechanics for free energy estimations in condensed-phase systems.
In the limit of large data or high energy evaluation budgets, all three investigated methods provide highly accurate absolute free energies as well as free energy differences between different crystalline phases.
While the continuous models outperform the discrete coupling flows for small  training budgets, inference times might still be slow and become a bottleneck for large system sizes.
Making these models size- and system-transferable will be decisive to establish them as preferable alternatives to traditional free energy estimators.

\section*{Supplementary Material}
The supplementary material contains details regarding the setup of the materials systems, the architecture of the machine learning models, a detailed discussion regarding the effective sampling size in CNFs and FEAT, the connection between the control term and score in FEAT, as well as details on the discretization of the path RND.

\begin{acknowledgments}
Dedicated to Christoph Dellago on the occasion of his 60th birthday. Dear Christoph, we warmly congratulate you and thank you for your inspiring contributions to molecular simulation and statistical Physics.
MS acknowledges financial support from Deutsche Forschungsgemeinschaft (DFG) through grant CRC 1114  ``Scaling Cascades in Complex Systems'', Project Number 235221301, Project B08  ``Multiscale Boltzmann Generators''. The Flatiron Institute is a division of the Simons Foundation.
JH acknowledges support from the University of Cambridge Harding Distinguished Postgraduate Scholars Programme. YD acknowledges support from Cornell University.
\end{acknowledgments}

\section*{Author declarations}

\subsection*{Conflict of Interest}
The authors have no conflicts to disclose.

\subsection*{Author Contributions}
{\bf Maximilian Schebek}:
Conceptualization (equal); Formal Analysis (equal); Investigation (equal); Software (equal); Visualization (lead); Writing -- original draft (lead); Writing -- review \& editing (equal).
{\bf Jiajun He}:
Conceptualization (supporting); Investigation (equal); Software (equal); Visualization (equal); Writing -- original draft (supporting); Writing -- review \& editing (equal).
{\bf Emil Hoffmann}:
Investigation (equal); Software (equal); Writing -- review \& editing (supporting).
{\bf Yuanqi Du}:
Conceptualization (equal); Investigation (supporting); Software (supporting); Visualization (equal); Writing -- original draft (supporting); Writing -- review \& editing (equal).
{\bf Frank No{\'e}}:
Writing -- review \& editing (supporting).
{\bf Jutta Rogal}:
Conceptualization (equal); Project administration (equal); Supervision (equal); Writing -- original draft (equal); Writing -- review \& editing (equal).

\section*{Data availability}
The models used in this work are available at 
\begin{itemize}
    \item \url{https://github.com/maxschebek/bgmat} (Coupling flows).
    \item \url{https://github.com/emil-ho/bg-cnf-fm-condensed-matter} (CNFs).
    \item\url{https://github.com/jiajunhe98/FEAT-condensed-matter}
 (FEAT).\\ 
\end{itemize}
All MD simulations were performed using the OpenMM package.~\cite{Eastman2023}

\newpage

\bibliography{article}

\setcounter{table}{0}
\setcounter{figure}{0}
\renewcommand{\thefigure}{S\arabic{figure}}
\renewcommand{\theequation}{S\arabic{equation}}
\renewcommand{\thetable}{S\arabic{table}}
\setcounter{equation}{0}
\setcounter{table}{0}
\setcounter{section}{0}
\renewcommand\thesection{\Alph{section}}
\renewcommand\thesubsection{\thesection.\arabic{subsection}}

\newpage
\section*{Supplementary Information}

\section{Systems}

\subsection{Lennard-Jones}

The pairwise LJ potential is given by~\cite{Frenkel2001-yl}
\begin{equation}
    u(r) = 4\epsilon \left[ {\left(  \frac{\sigma}{r} \right)}^{12}  - {\left(  \frac{\sigma}{r} \right)}^{6} \right],
\end{equation}
where $r$ is the two-particle distance. $\varepsilon$ and $\sigma$   define the characteristic length and energy scales, respectively. A cutoff radius $r_{\rm cut}$ is typically employed and the potential is shifted to be continuous at the cutoff. This results in the following interaction potential: 
\begin{equation}
    u_\text{cut}(r) = 
    \begin{cases}
        u(r)   - u(r_\text{cut})          & \text{if  }  r \leq r_\text{cut}, \\
        0                                                   & \text{else}.
\end{cases}
\end{equation}
To simplify comparisons between different systems, reduced units based on $\varepsilon$ and $\sigma$ are commonly used, and quantities in these units are typically denoted by an asterisk. Cutoffs of $r_{\rm cut}=2.7^*$ (FCC with $N=256$) and $r_{\rm cut}=2.2^*$ (FCC and HCP with $N=180$) were employed. 

\subsection{Monoatomic water}
Monoatomic water (mW)~\cite{molinero2009water} is based on the Stillinger-Weber potential~\cite{stillinger1985computer} which features two-body ($\phi_2$) and three-body  ($\phi_3$) interactions, of which the latter enforce tetrahedral coordinations. The total potential energy is given by
\begin{equation}
    U_\text{SW}(\mathbf{x}) = \sum_i \sum_{j > i}  \phi_2(d_{ij}) +  \lambda_3 \sum_i \sum_{j \neq i} \sum_{k > j} \phi_3(d_{ij}, d_{ik}, \theta_{ijk})\quad,
\end{equation}
with
\begin{align}
  \phi_2(r) & = A \epsilon \left[ B   {\left(  \frac{\sigma}{r} \right)}^{4} - 1 \right] \exp\left(\frac{\sigma}{r - a \sigma}\right)\quad ,\\
    \phi_3(r,s, \theta) & = \lambda \varepsilon {\left( \cos\theta - \cos\theta_0  \right)}^2 \exp{ \left(\frac{\gamma \sigma}{r - a \sigma}\right)} \exp{\left(\frac{\gamma \sigma}{s - a \sigma}\right)},
\end{align}
where $d_{ij}$ denotes a two-particle distance and $\theta_{ijk}$ is the angle formed by the three atoms. All parameters are fixed except for $\varepsilon$, $\sigma$, and $\lambda_3$, which are adjusted to model a specific system. Here, $\lambda_3$ sets the strength of the three-body interactions, while $\varepsilon$ and $\sigma$ define the characteristic energy and length scales. The common constants are $A = 7.049556277$, $B = 0.6022245584$, $a = 1.8$, $\theta = 109.47^\circ$, and $\gamma = 1.2$. For the monatomic water (mW) model, the tuned values are $\lambda = 23.15$, $\varepsilon = 6.189~\text{kcal/mol}$, and $\sigma = 2.3925~\text{\AA}$.~\cite{molinero2009water}  

\subsection{Computational settings}
Cubic and hexagonal mW ice were modeled at \(T = 200\) K at a density of $\rho=1.004$ gcm$^{-3}$. The LJ crystals were modeled at $T^*=2.0$ and $\rho^*=1.28$.\\

\subsection{Absolute free energies}
Reference values for absolute values of the reduced free energies are given in Tab.~\ref{tab:fe}.

\begin{table*}[h]
\caption{\label{tab:fe} 
Reference reduced free energies $\beta F/N$ (partially from~\cite{Wirnsberger_2022}).}
\begin{tabular*}{\textwidth}{@{\extracolsep{\fill}} l l r @{}}\hline
System  & $N$  & MD+MBAR \\ 
\hline
LJ      & 256 FCC ($r^*_{\rm cut} = 2.7$) & 3.10798(9) \\
LJ      & 500 FCC ($r^*_{\rm cut} = 2.7$) & 3.12262(10) \\ 
LJ      & 108 FCC ($r^*_{\rm cut} = 2.0$) & 3.8333(4) \\ 
LJ      & 180 FCC ($r^*_{\rm cut} = 2.2$) & 3.5346(3) \\ 
LJ      & 180 HCP ($r^*_{\rm cut} = 2.2$) & 3.5293(1) \\ 
\hline
Ice Ic  & 64  & -25.16306(20) \\
Ice Ic  & 216 & -25.08234(5) \\
Ice Ic  & 512 & -25.06156(3) \\
\hline
Ice Ih  & 64  & -25.18687(26) \\
Ice Ih  & 216 & -25.08975(14) \\
Ice Ih  & 512 & -25.06480(4) \\
\hline
\end{tabular*}
\end{table*}

\subsection{Convergence MBAR}
\begin{figure}[h]
    \centering
    \includegraphics{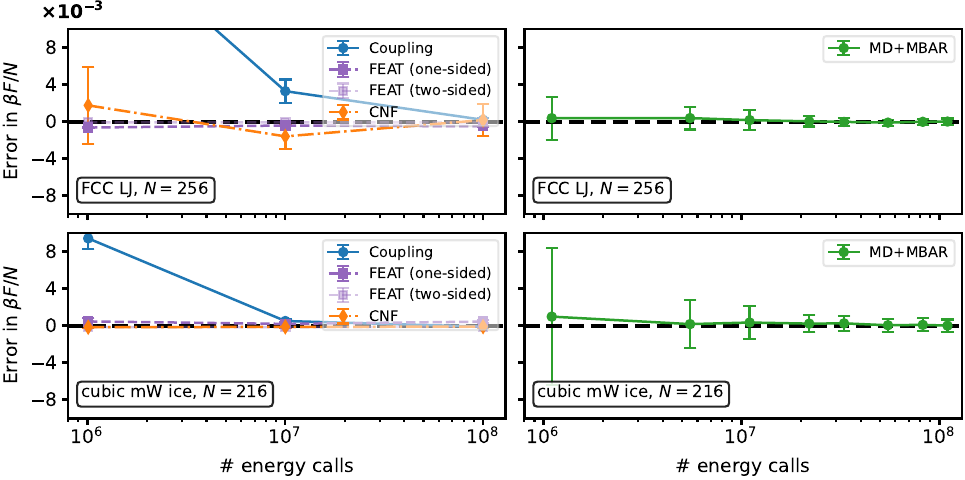}
    \caption{Convergence of the absolute free energy estimate for FCC LJ with $N=256$ (upper row) and cubic mW ice $N=216$ (lower row) with respect to the number of energy evaluations for the generative models and MD+MBAR. Generative-model estimates were obtained from 10k inference samples at three training budgets (low, medium, high). For MD+MBAR, the total cost includes energy evaluations used for MD sampling as well as those required for MBAR evaluation of the free energy difference.}
    \label{fig:convergence_mbar}
\end{figure}

Fig.~\ref{fig:convergence_mbar} presents a detailed comparison of the computational cost required to obtain converged absolute free energy estimates using generative models and conventional MD+MBAR. We report the error in the free energy as a function of the total number of energy evaluations, which serves as a hardware-independent measure of computational effort. For the generative models, the cost includes training and inference evaluations, with estimates computed from 10k generated samples at three training budgets. For MD+MBAR, the cost accounts for both MD sampling and the subsequent MBAR analysis. For the MBAR calculations, we used 100 intermediate stages and varied the number of samples collected per stage. The largest setting corresponds to 1k samples per stage, resulting in a total of $100 \times 10^3 \times 10^3 = 10^8$ energy evaluations, assuming that $10^3$ MD steps are required between subsequent samples for decorrelation. An additional, comparatively small overhead arises from evaluating all samples in all intermediate potentials in the MBAR analysis.

\section{Model architectures}

\subsection{Prior distribution}
The Gaussian prior distribution was chosen to be identical across all models, with zero mean and a standard deviations of 0.2\,\AA~for the mW systems and to 0.05\,$\sigma$ for the LJ systems. To represent the Einstein crystal, the displacements were added to the corresponding reference lattice of the various structures.

\subsection{Coupling flows}
We rely on the augmented coupling flow architecture introduced in Ref.~\onlinecite{schebek2024efficient}. Augmented flows  introduce auxiliary variables $\ba\in\mathbb{R}^{3N}$   and enable the splitting to be performed between physical and auxiliary variables, which retains full three-dimensional coordinates within the physical and auxiliary space.
In our architecture, we model the displacements from the ideal crystal lattice $\mathbf{L}^0$ rather than absolute positions, and the joint prior distribution is defined as 
\begin{equation}\quad\label{eq:joint_base}
    q(\bx,\ba) =q(\bx)\,\mathcal{N}(\ba;\bx, \eta^2\mathbf{I}) \quad.
\end{equation}
The position of each particle $i$ is updated as
\[
\bx_i' = g(\bx_i \mid \bh^\ba_i),
\]
where $g$ is a bijector parametrized by the embedding $\bh^\ba_i$.  
 The particle embeddings are computed using a graph neural network (GNN) composed of $L$ layers, with updates at layer $l$ given by
\begin{equation}\label{eq:gnn}
\begin{aligned}
\bd_{ij} &= {\rm sinusoidal}([ [\ba_i + \mathbf{L}_i^0]_{\rm PBC} - [\ba_j+\mathbf{L}_i^j ]_{\rm PBC}\,]_{\rm PBC}; \boldsymbol{\omega}_d)\quad , \\[1mm]
\bm^l_{ij} &= \phi_e(\bh_i^l,\bh_j^l,\mathbf{d}_{ij})\quad , \\[1mm]
\bm^l_i &= \sum_{j\in\mathcal{N}_i} \bm^l_{ij}\quad , \\[1mm]
\bh_i^{\ba,l+1} &= \phi_h(\bh_i^{\ba,l},\bm^l_i)\quad ,
\end{aligned}
\end{equation}
where $\mathcal{N}_i$ denotes the set of neighbors of particle $i$, and $\phi_e$ and $\phi_h$ are learnable functions. The notation $[]_{\rm PBC}$ indicates that sums and differences of vectors are to be taken with respect to the periodic boundary conditions. The bijector $g$ is implemented as rational quadratic splines.~\cite{durkan2019neural} The generated augmented distribution can be obtained via the change-of-variable theorem as
\begin{equation}\label{eq:cov_augmented}
    q'(\bx', \ba') = q(\bx,\ba) |\det J_{f_\theta}(\bx,\ba)|^{-1}\quad ,
\end{equation}
while the marginal of the generated distribution can be evaluated according to 
 \begin{equation}\label{eq:marginal}
        q'(\bx') \approx  \frac{1}{M} \sum_{m=1}^M \frac{q'(\bx',\ba_m)}{\pi(\ba_m|\bx')}\quad,
    \end{equation}
which requires $M$ additional inverse passes per sample.

\begin{figure}
    \centering
    \includegraphics{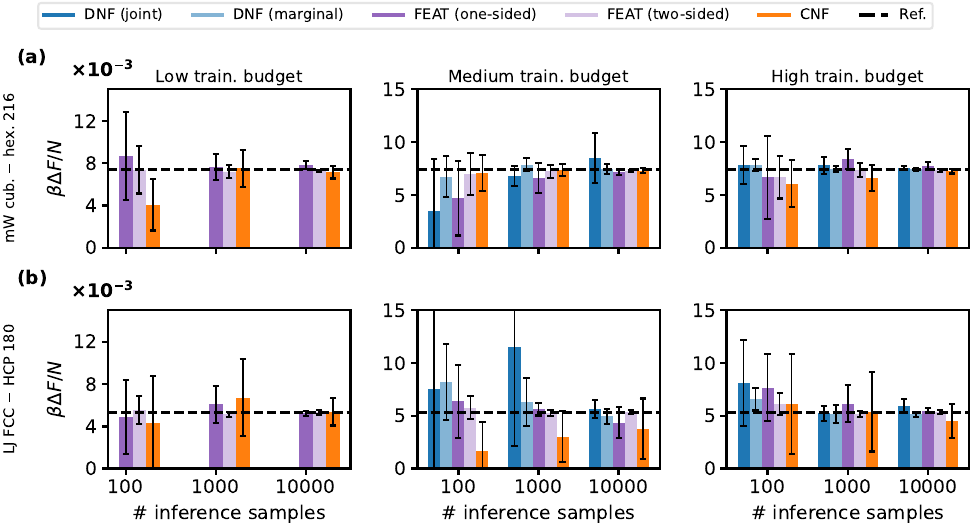}
    \caption{Figure showing the same results as Fig.~4 in the main text, now including free energy estimates obtained from the marginal formulation of the coupling flows, where the auxiliary degrees of freedom are integrated out according to Eq.~\eqref{eq:marginal} with $M=200$.}
    \label{fig:marginal}
\end{figure}

For the maximum training budget of $10^8$ energy evaluations, training took about three days on four NVIDIA A5000 GPUs for the larger systems. Further details on the architecture as well hyperparameter choices can be found in Ref.~\onlinecite{schebek2024efficient}. Reported free energy estimates and sampling efficiencies in the main text were evaluated over the augmented space. Figure~\ref{fig:marginal} shows the corresponding results when the marginal is evaluated, i.e. when the auxliary degrees of freedom are integrated out according to Eq.~\eqref{eq:marginal} setting $M=200$. Since the marginal ESS is always higher than the joint one,~\cite{schebek2025scalable} evaluating the marginal can reduce the variance of the free energy estimates, as can be seen in the figure for the medium and high training budgets which allows to reduce the number of required energy evaluations during inference.

\subsection{Continuous Normalizing Flow and FEAT}
We adopt the equivariant graph neural network (EGNN) parameterization of Ref.~\onlinecite{satorras2021equivariant} for the vector field and adapt it to periodic boundary conditions.

The vector field operates on absolute positions as input, which lie on the flat torus manifold. Stochastic interpolants were generalized to Riemannian Manifolds in Ref.\onlinecite{chen2024flow}, which enables the direct use of absolute positions for the CNF models. 
FEAT is, however, not formulated on arbitrary manifolds, and, similar to the coupling flows, the displacements of the atoms from the ideal crystal lattice are used, with the equilibrium lattice added onto them for the distance calculations in the EGNN. 
More precisely, the states $\bx_A$ and  $\bx_B$, and all $\bx_t$ are defined as the displacements for FEAT, while the control term is calculated with the same EGNN as used in the CNFs by first converting the displacements back into absolute positions by adding the equilibrium lattice, i.e., $b^\theta(t, \bx_t) = \text{EGNN}(t, [\bx_t + \bx^{\text{eq}}]_{\rm PBC})$.
We choose to use FEAT directly on displacements as we want to ensure that the prior distribution corresponds to a  simple Gaussian without PBC, so that the score can be analytically derived from the learned vector field.
Otherwise, both networks are identical.

The time $t$ is embedded with an exponential-frequency sinusoidal embedding $\gamma(t)$, which is broadcast to every node. We concatenate additional sinusoidal features $\psi(\mathbf{x}_i^{\mathrm{eq}})$ computed from the equilibrium position. The resulting node features are $\mathbf{h}_i^{0} = [\gamma(t),\psi(\mathbf{x}_i^{\mathrm{eq}})]$. The vector field $v_\theta(t,\mathbf{x})$ is then modeled by a periodic EGNN with $L$ message-passing layers, producing updated coordinates $\{\mathbf{x}_i^{L}\}$, and we return the tangent-space velocity as the displacement $\mathbf{x}^{L}-\mathbf{x}^{0}$.

Concretely, at layer $l$ we define the PBC displacement and distance
\begin{equation}
\mathbf{r}_{ij}^{\,l} = [\mathbf{x}_i^{\,l}-\mathbf{x}_j^{\,l}]_{\mathrm{PBC}}
\qquad
d_{ij}^{\,l}=\|\mathbf{r}_{ij}^{\,l}\|,
\end{equation}
and use a Gaussian smearing radial expansion $\rho(d_{ij}^{\,l})$. In addition, we provide a scalar edge attribute $(d_{ij}^{\ 0})^2$ computed once from the initial configuration.

The layer updates implemented by the network are
\begin{align}
\mathbf{z}_{ij}^{\,l} &= \phi_e\left(\mathbf{h}_i^{\,l}, \mathbf{h}_j^{\,l}, \rho(d_{ij}^{\,l}), e_{ij}\right), \\[1mm]
\mathbf{m}_i^{\,l} &= \sum_{j\in\mathcal{N}(i)} \mathbf{z}_{ij}^{\,l}, \\[1mm]
\mathbf{h}_i^{\,l+1} &= \phi_h\left([\mathbf{h}_i^{\,l}, \mathbf{m}_i^{\,l}]\right), \\[1mm]
\mathbf{x}_i^{\,l+1} &= \mathbf{x}_i^{\,l} + \sum_{j\in\mathcal{N}(i)} 
\frac{\mathbf{r}_{ij}^{\,l}}{d_{ij}^{\,l}+1}
\cdot \phi_x(\mathbf{z}_{ij}^{\,l}),
\end{align}
where $\phi_e$ and $\phi_h$ are MLPs, and $\phi_x$ is an MLP ending in a single scalar output (applied per edge) that scales the normalized displacement direction. Finally, the CNF vector field returned by the dynamics model is the layer-wise coordinate change
\begin{equation}
v_\theta(t,\mathbf{x}^0) = \mathbf{x}^{L}-\mathbf{x}^{0}.
\end{equation}
Hyperparameters are listed in table~\ref{tab:vecfield_hyperparams}.

\begin{table}[]
\caption{Hyperparameters of the vector field parametrization used by the CNF and the FEAT estimator.}
\label{tab:vecfield_hyperparams}
\begin{tabular}{lc}
\hline
Number of message passing layers         & 10               \\
Hidden layer width                       & 32               \\
Number of time embedding features        & 10               \\
Number of distance expansion features    & 10               \\
Number of equilibrium positions features & 5                \\
Total parameters                         & $\approx 80,000$ \\ \hline
\end{tabular}
\end{table}

Training for $10^6$ steps takes between 15 and 30 hours on a single A5000 GPU, depending on the system size.
For the CNF, generating samples and computing exact density change with the exact trace of the Jacobian becomes prohibitively expensive, due to its quadratic scaling. Hence, we estimate the trace of the Jacobian with Hutchinson's estimator.~\cite{Hutchinson01011990}. While the divergence estimates are unbiased, a noisy estimate of the density change will lead to a biased estimate for the free energy difference. Hence, a rather large number of Hutchinson probes is needed, with each probe costing one forward pass, which is still orders of magnitude faster than the exact evaluation of the trace of the Jacobian. Additionally, a discretization error also incurs a bias in the free energy difference.
With 50 probes per step, the reduced free energy estimates converged within around $5\cdot10^-4$~\cite{Hoffmann2026Boltzmann}. Similarly, 50 integration steps using \textit{rk4} (200 function evaluations) are sufficient to converge the estimates.
It takes about half an hour on a single A5000 GPU to generate 1,000 samples for the biggest system.

For FEAT, we calculate the work using Eq.~\eqref{eq:work_gaussian}. This formulation does not require the evaluation of the trace of the Jacobian, and hence is much faster than CNF.
We discretize the SDE with 500 steps, and it only takes ~1-2 minutes to generate 1,000 samples for the biggest system.
One caveat for FEAT is that the score can becomes highly inaccurate for $t=0, 1$.
This does not influence the correctness of the estimator, but it can lead to lower ESS (higher variance of the work).
In our implementation, we therefore directly use the exact forces  $\nabla u_A(\bx_0)$ or $\nabla u_B(\bx_1)$ for $t=0$ or $1$.
This will not increase the number of target energy evaluations, as we need to evaluate the energies $u_A(\bx_0)$ and $u_B(\bx_1)$ anyway when calculating the work.
The diffusion coefficient $\sigma_t$ for the SDE is set to $0.01$ for all $t$ when using \AA~ as the unit of length.

All CNF models were trained for 1M steps with a learning rate of 0.0001 and the schedule-free implementation of the AdamW optimizer~\cite{loshchilov2018decoupled, defazio2024the}.
A batch size of 128 was used together with mini-batch optimal transport reordering.~\cite{tong2023improving,pmlr-v202-pooladian23a}. Since the reordering is carried out by the CPU-based data-loading pipeline and overlaps with GPU computation, it does not affect overall training time.
For FEAT, we train with Adam with a learning rate of 0.0001 and also keep an EMA with rate 0.999 following Ref.~\onlinecite{he2025feat}.
As the analytical relation between score and control requires independent pair between $p_A$ and $p_B$, we did not use any form of OT plan during training.

\section{Effective sampling size}
The Kish effective sample size (ESS)~\cite{Kish1965-es} provides a measure of the number of statistically independent samples represented in a weighted ensemble. It accounts for the reduction in statistical efficiency caused by unequal sample weights and is commonly used to assess the reliability of reweighted distributions. For a set of normalized weights \( \{w_i\} \), the Kish ESS is defined as
\begin{align}
    N_{\mathrm{eff}} = \frac{\left(\sum_i w_i\right)^2}{\sum_i w_i^2}\quad,
\end{align}
which reduces to the total number of samples when all weights are equal and decreases as weight heterogeneity increases.

\section{Detailed Discussion on Trade-off between FEAT and CNF}

Except for the LJ system with 180 particles, CNF consistently presents higher ESS than FEAT. This  behavior is, however, expected: FEAT effectively performs importance sampling over the path space of the SDE, while CNF directly performs importance sampling in the target (marginal) space.
Let $\tilde{p}_{A'}^{\text{FEAT}}$ and $\tilde{p}_{A'}^{\text{CNF}}$ denote the pushforward measures from $A$ induced by the learned transport in FEAT and CNF, respectively. Assume that FEAT and CNF achieve the same quality of transport, i.e., the $f$-divergence 
$D_f[\tilde{p}_{A'}^{\text{FEAT}}\| p_B] = D_f[\tilde{p}_{A'}^{\text{CNF}}\| p_B]$.
In this case, the ESS of FEAT is still lower than that of CNF, since
$D_f[\overrightarrow{\mathbb{P}}^{\,b}\| \overleftarrow{\mathbb{P}}^{\,b}]
\geq
D_f[\tilde{p}_{A'}^{\text{FEAT}}\| p_B]
=
D_f[\tilde{p}_{A'}^{\text{CNF}}\| p_B]$.
The first inequality follows from the data processing inequality, since marginalization from path space to the terminal distribution cannot increase the divergence.

However, while CNF presents higher ESS, this improvement does not come without a price.
First, CNF is extremely slow compared to FEAT due to the requirement of divergence calculation. Even when using Hutchinson’s estimator and a small number of integration steps, CNF is still $10$--$30\times$ more expensive than FEAT in our experiments.
Second, FEAT is asymptotically unbiased even in the presence of discretization error. This is because the work defined in Eq.~\eqref{eq:work_gaussian} already takes the discretization error into account and ensures that the entire estimator remains consistent. A more formal proposition can be found in Proposition~3.3 of Ref.~\onlinecite{he2025feat}.
In contrast, CNF is strictly unbiased only in continuous time using the exact Jacobian trace, while in practice trace approximations and discretization errors introduce a bias that will not disappear by using more samples. Therefore, in summary, FEAT and CNF can be viewed as a trade-off between variance and bias, as well as computational efficiency.

\section{Connection between Score and Vector Field}

In this section, we derive the connection between score and vector field.~\cite{albergo2025stochastic}
In our experiments, we choose $\bx_t  = I_t  = (1-t) \bx_0 + t \bx_1$.
Therefore, the optimal control term is given by
\begin{align}
    b_t^*(\bx_t) &=  \mathbb{E}[\bx_1|\bx_t] -  \mathbb{E}[\bx_0|\bx_t]\\
     &=  \mathbb{E}[\bx_1|\bx_t] -  \frac{1}{1-t} \mathbb{E}[\bx_t - t \bx_1|\bx_t]\\
     &=  \frac{1}{1-t}\mathbb{E}[\bx_1|\bx_t] -  \frac{\bx_t}{1-t} 
\end{align}
Assume $\bx_0 \sim \mathcal{N}(0, v\mathrm{Id})$ with $v$ as the variance.
In DSM, the optimal score is given by
\begin{align}
    s^*_t(\bx_t) = \frac{t\mathbb{E}[\bx_1 |\bx_t]- \bx_t}{(1-t)^2 v}
\end{align}
Therefore, we have 
\begin{align}
    s^*_t(\bx_t) &= \frac{t  (1-t)b_t^* (\bx_t)  + t \bx_t  - \bx_t}{(1-t)^2 v}\\
    &= \frac{t   b_t^* (\bx_t) - \bx_t }{(1-t) v}
\end{align}

\section{It\^o's Integral and Path RND}
For FEAT, we calculate the work defined in Eq.~\eqref{eq:work_fb_rnd} using $R$ given by Eq.~\eqref{eq:ito}.
In this section, we provide a detailed background and explanation on the calculation of the It\^o's Integral.

Consider a discretized trajectory $[\bx_{t_0}, \cdots, \bx_{t_{K}}]$ as solution to the SDE in Eq.~\eqref{eq:fwd_sde} (or a solution to Eq.~\eqref{eq:bwd_sde}).
For $a_t(\bx_t)$ with mild boundedness condition, the (forward) It\^o integral is defined by
\begin{align}
\int_0^1 a_t(\bx_t)\cdot \mathrm{d} \overrightarrow{\bx_t}
:= \lim_{\|\mathcal{P}\|\to 0}\sum_{k=0}^{K-1}
a_{t_k}(\bx_{t_k})\big( \bx_{t_{k+1}}-\bx_{t_k}\big),
\end{align}
where $\mathcal{P}=\{0=t_0<\cdots<t_K=1\}$ is the partition of the time horizon $[0, 1]$ and $\|\mathcal{P}\|=\max_k (t_{k+1}-t_k)$.
\par
Similarly, the backward It\^o integral is defined by
\begin{align}
\int_0^1 a_t(\bx_t)\cdot  \mathrm{d} \overleftarrow{\bx_t}
:= \lim_{\|\mathcal{P}\|\to 0}\sum_{k=0}^{K-1}
a_{t_{k+1}}(\bx_{t_{k+1}})\big( \bx_{t_{k+1}}-\bx_{t_k}\big),
\end{align}
Note that \emph{the direction of the It\^o's integral does not depend on the direction of the SDE}.
The backward It\^o integral can be evaluated on a solution to forward SDE, while the forward integral can also be evaluated on backward SDE's solution.
Therefore, \emph{$R$ defined in Eq.~\eqref{eq:ito} can be directly applied to calculate the work associated with both forward and backward trajectories}. 
If we only simulate the forward SDE, calculate its work and estimate the free energy using the escorted Jarzynski equality (Eq.~\eqref{eq:esc_jarzinsky}), we obtain the free-energy estimator which we call one-sided FEAT in our experiments; if we simulate both forward and backward SDEs, calculate their work and estimate the free energy using Eq.~\eqref{eq:two-sided-F}, we obtain the two-sided FEAT estimator in our experiment.

An alternative and more intuitive way to view the forward-backward stochastic integral is looking at the discrete counterpart with the Euler-Maruyama (EM) method.
We discretize the forward and backward SDE using EM  with $ \{0=t_0<\cdots<t_K=1\}$ and $\Delta t= t_{k+1}-t_k$:
\begin{align}
& p(\bx_{t_{k+1}}|\bx_{t_k}) = \mathcal{N}(\bx_{t_{k+1}}| \bx_{t_k} +   b_{t_k}^\theta(\bx_{t_k})\Delta t + \sigma_{t_k}^2 s^\phi_{t_k}(\bx_{t_k})\Delta t,  2\sigma_{t_k}^2 \Delta t) \label{eq:fwd_sde_gaussian}\\
& p(\bx_{t_{k-1}}|\bx_{t_k}) = \mathcal{N}(\bx_{t_{k-1}}| \bx_{t_k} -   b_{t_k}^\theta(\bx_{t_k})\Delta t + \sigma_{t_k}^2 s^\phi_{t_k}(\bx_{t_k})\Delta t,  2\sigma_{t_k}^2 \Delta t) \quad, \label{eq:bwd_sde_gaussian}
\end{align}
then
\begin{align}\label{eq:work_gaussian}
   \hat{R} =  - \ln \frac{\prod_{k=1}^{K} p(\bx_{t_{k-1}}|\bx_{t_k}) }{ \prod_{k=0}^{K-1} p(\bx_{t_{k+1}}|\bx_{t_k}) } \quad ,
\end{align}
with $\lim_{\Delta t \rightarrow 0 }\hat{R} = R$. 
Eq.~\eqref{eq:work_gaussian} can be evaluated on any trajectory $\{ x_{t_k}\}_{k=0}^K$, independent of if the trajectory is generated from the forward or backward SDE.
In our experiments, we directly use Eq.~\eqref{eq:work_gaussian} to calculate the work numerically.
\par
With this expression, the one-sided FEAT estimator works as follows:
\begin{enumerate}
    \item \emph{simulation.} Starting from $\bx_0\sim p_A$, we draw $\bx_{t_1}, \bx_{t_2}, \cdots, \bx_{t_K=1}$ sequentially with Eq.~\eqref{eq:fwd_sde_gaussian}.;
    \item \emph{work evaluation.} Calculate work for each sample trajectory with $\widetilde{w}=u_B({\bx_1}) - u_A({\bx_0}) + \hat{R}$, $\hat{R}$ calculated by Eq.~\eqref{eq:work_gaussian} which requires to evaluate both Eq.~\eqref{eq:fwd_sde_gaussian} and Eq.~\eqref{eq:bwd_sde_gaussian} along the trajectories.
    \item \emph{Free energy estimation.} Estimate the free energy with the escorted Jarzynski equality in Eq.~\eqref{eq:esc_jarzinsky}.
\end{enumerate}
The two-sided FEAT estimator also performs the \emph{simulation} and \emph{work evaluation} step for a forward trajectory. In addition, it repeat the process for the backward process:
\begin{enumerate}
    \item \emph{backward simulation.} Starting from $\bx_1\sim p_B$, we draw $\bx_{t_{K-1}}, \bx_{t_{K-2}}, \cdots, \bx_{t_0=0}$ sequentially with Eq.~\eqref{eq:bwd_sde_gaussian};
    \item \emph{backward work evaluation.} Calculate work for each sample trajectory with $\widetilde{w}=u_B({\bx_1}) - u_A({\bx_0}) + \hat{R}$, $\hat{R}$ calculated by Eq.~\eqref{eq:work_gaussian} which requires the evaluation of both Eq.~\eqref{eq:fwd_sde_gaussian} and Eq.~\eqref{eq:bwd_sde_gaussian} along the backward trajectories. Note that this expression does not change compared to the forward simulation.
    \item \emph{Free energy estimation.} Estimate the free energy with the minimal-variance estimator in Eq.~\eqref{eq:two-sided-F} using both the work from the forward and backward trajectories.
\end{enumerate}

\end{document}